\documentclass[letterpaper,11pt]{article}
\pdfoutput=1

\usepackage{jheppub}
\usepackage{multirow}

\usepackage{subcaption}
\usepackage[countmax]{subfloat}
\usepackage{amssymb}
\usepackage{amsmath}
\usepackage{color}
\usepackage{graphicx}
\usepackage{verbatim}
\usepackage{amsthm}
\usepackage{slashed}
\usepackage{hyperref}

\preprint{}

\title{
Top-quark mass determination from $t$-channel single top production at the LHC
}

\author{Mei Sen Gao$^{1}$, Shu Run Yuan$^{1}$, Jun Gao$^{1,\,2,\,3}$}
\affiliation{
$^1$INPAC, Shanghai Key Laboratory for Particle Physics and Cosmology, School of
Physics and Astronomy, Shanghai Jiao Tong University, Shanghai 200240, China
}
\affiliation{
$^2$Key Laboratory for Particle Astrophysics and Cosmology, Shanghai 200240, China
}
\affiliation{
$^3$Center for High Energy Physics, Peking University, Beijing 100871, China
}

\emailAdd{jung49@sjtu.edu.cn}

\abstract{
We study the determination of the top-quark mass
using leptonic observables in $t$-channel single top-quark production
at the LHC.
We demonstrate sensitivity of transverse momentum of the charged
lepton on the input top-quark mass.
We present predictions at next-to-next-to-leading order (NNLO) in QCD with narrow width approximation
and structure function approach.
Further corrections due to parton shower and hadronization, non-resonant
and non-factorized contributions are discussed.
To reduce impact of SM backgrounds we propose to use the charge weighted
distribution for the measurement, i.e., differences between distributions
of charged lepton with positive and negative charges.
By modeling both signal and background processes, we found the projections for (HL-)LHC
to be promising, with a total
theoretical uncertainty on the extracted top-quark mass of about $1\sim 2$~GeV.
}
  

\begin{document} 

\maketitle

\section{Introduction}

The top quark ($t$) is the heaviest particle in the standard model~(SM).   
The mass of top quark has been one of the most important input parameters
of the SM and its experimental determination is crucial for precision test
of the SM.
For example, the recent global analysis of electroweak precision
observables reveals a good agreement with 
top-quark mass from direct measurements~\cite{1803.01853}.
Top quark also plays important role in renormalization group running of
the SM especially due to the large Yukawa coupling,
where the stability of the electroweak vacuum~\cite{hep-ph/0104016} is
sensible to the precise top quark mass.

The mass of top-quark can be measured directly at Tevatron or LHC in top-quark
pair production with subsequent decays, e.g., through invariant mass
distributions of various decay products, with which the CDF, D0, ATLAS and CMS
collaborations have reported an unprecedented precision of about
0.5~GeV~\cite{1407.2682,1810.01772,1812.10534}.
The above measurements are supposed to be affected
by various non-perturbative QCD effects that are modeled by
MC event generators.
The associated systematic uncertainties become more
and more important as the experimental precision improves
and have been studied extensively~\cite{1412.3649,1712.02796,
1807.06617,1902.05035,FerrarioRavasio:2019vmq}.
There are also discussions on intrinsic ambiguities of a precise
definition of the top-quark mass due to infrared renormalon effects~\cite{1605.03609,1704.01580,1706.08526}. 
Many alternative methods on determination of the top-quark mass
have been proposed and carried out at the LHC in order to scrutinize the
experimental precision.
That includes utilizing kinematic variables other than conventional
invariant masses in top-quark pair production~\cite{hep-ph/9906349,1407.2763,1603.03445,1603.06536,1608.03560,1709.09407}, 
and using measurements of total inclusive cross sections~\cite{1406.5375,1603.02303,1812.10505} or
of distributions inclusive with respect to decay products of top quark~\cite{1904.05237,1908.02179,2004.03088}.
There are cases of using processes of associated production of
top-quark pair with a jet~\cite{1303.6415} or single top-quark
production~\cite{ATLAS:2014baa,1703.02530,1608.05212}. 
Besides, at future electron-positron colliders the top-quark mass
can be measured much more precisely via an energy scan at threshold of top-quark
pair production~\cite{1506.06864}.
A recent review on various topics in determination of top-quark mass
can be found in~\cite{2004.12915}.

In this work we perform a theoretical study on determination
of the top-quark mass
through $t$-channel single top-quark production at the LHC.
In particular we study in details transverse momentum distributions of the
charged lepton from decay of the top quark with which we determine the top-quark mass.
Similar approaches have been adopted for mass measurements in top-quark pair production~\cite{1407.2763}.
It has been advertised that methods with pure leptonic variables, i.e., not directly
involving jets, will be less affected by various non-perturbative
effects, as well as by uncertainties from jet energy scale.
Measuring top-quark mass in single top-quark production~\cite{ATLAS:2014baa,
1703.02530,1608.05212} is complementary
to those measurements in pair production because of different dynamics of production
including QCD color flows, which leads to different theoretical uncertainties.
It can provide independent and valuable inputs to the global determination
of top-quark mass.
Besides, significant efforts have been made to improve the theoretical
description of $t$-channel single top-quark production.
We note that QCD corrections in single top-quark productions are
in general much smaller than those in pair production.
The next-to-leading order (NLO) QCD corrections in the 5-flavor scheme (5FS) are calculated 
in Refs.~\cite{NUPHA.B435.23,hep-ph/9603265,hep-ph/9705398,hep-ph/9807340,hep-ph/0102126,
hep-ph/0207055,Sullivan:2004ie,hep-ph/0408158,hep-ph/0510224,hep-ph/0504230,1007.0893,
1012.5132,1102.5267,1305.7088,1406.4403,Carrazza:2018mix}.
The NLO calculation in the 4-flavor scheme (4FS) is carried out in Ref.~\cite{0903.0005}.
Full NLO corrections including top-quark leptonic decay are studied within  
the on-shell top-quark approximation~\cite{hep-ph/0408158,hep-ph/0504230,1012.5132}
and beyond~\cite{1102.5267,1305.7088,1603.01178,Neumann:2019kvk}.
Code for fast numerical evaluation at NLO is 
provided in Ref.~\cite{1406.4403}.
The NLO electroweak corrections are also calculated~\cite{1907.12586}. 
Soft gluon resummation is considered in
Refs.~\cite{1010.4509,1103.2792,1210.7698,1510.06361,1801.09656,Kidonakis:2019nqa,Cao:2019uor}.
Matching NLO
calculations to parton shower is done in the framework of POWHEG and
MC@NLO~\cite{hep-ph/0512250,0907.4076,1207.5391,1603.01178}. 
Next-to-next-to-leading order (NNLO) QCD
corrections with a stable top quark are calculated in
Ref.~\cite{1404.7116}.
The study here are based on the NNLO calculations including top-quark
leptonic decay under narrow width approximation (NWA) as
developed in Refs.~\cite{Berger:2016oht,1708.09405,
1807.03835,2005.12936} that provide a realistic parton-level
simulation at NNLO.
Moreover, we have used the structure function approach~\cite{Lindfors:1985zz,Han:1992hr,Stelzer:1997ns},
where the $t$-channel production can be factorized into two
charged-current deep-inelastic scatterings with light quarks and heavy quarks
respectively.
Gluon exchanges between the two quark lines contribute at NNLO.
They are suppressed by QCD colors and neglected in our study.
Progresses on calculation of those corrections have been made
in~\cite{Assadsolimani:2014oga,Meyer:2016slj}.
However, we should point out several potential problems on determination
of top-quark mass via single top-quark production.
It suffers from large backgrounds due to top-quark pair production
as well as production of W/Z boson with jets.
A pure signal sample can only be obtained in a signal-enriched
fiducial volume as shown in the ATLAS and CMS
measurements~\cite{ATLAS:2014baa,1703.02530}.
That indicates the measurements are less inclusive and also have relatively
low statistics as compared to measurements in pair production.
The former one is less concerned as far as precise theory
predictions can be provided which is the main topic of this paper.
The shortcoming on statistics can also be overcome thanks to the high luminosity
of LHC.

The rest of our paper is organized as follows.
In Sec.~\ref{sec:lep}, we describe leptonic observables and the
sensitivity to top-quark mass.
In Sec.~\ref{sec:the}, we present our nominal predictions
on leptonic distributions including their intrinsic and parametric uncertainties.
Sec.~\ref{sec:alt} provides results of alternative theory
predictions including those from different heavy quark schemes and from MC generators.
In Sec.~\ref{sec:dis} we address further questions on both
theory and experimental sides related to the measurement and in Sec.~\ref{sec:pro}
we show our projected precision of measurements at (HL-)LHC.
Finally our summary and conclusions are presented in 
Sec.~\ref{sec:sum}.

\section{Leptonic observables}
\label{sec:lep}

We demonstrate dependence of leptonic variables on the top
quark mass in single top-quark production.
We use on-shell renormalization scheme in perturbative calculations.
Thus the top-quark mass we refer to in the remaining sections is always the
pole mass.  
Specifically we focus on transverse momentum distributions of the charged
lepton. 

We start with a pedagogical discussion based on a calculation at Born level.
Kinematic distributions of the charged-lepton can be understood as below.
For decay of an on-shell top quark in its rest frame, the
triple differential decay width can be expressed as~\cite{hep-ph/9402326}
\begin{equation}\label{lodecay}
\frac{d\Gamma}{dxdyd\cos \theta}=\frac{G_F^2m_t^5}{32\pi^3}
\frac{|V_{tb}|^2}{(1-y/\bar y)^2+\gamma^2}x(x_m-x)(1+S\cos\theta), 
\end{equation}
with
\begin{equation}
x_m=1-\epsilon^2,\,\,\epsilon=m_b/m_t,\,\,\bar y=m_W^2/m_t^2,\,\,
\gamma=\Gamma_W/m_W,
\end{equation}
where $m_b$, $m_t$, $m_W$ and $\Gamma_W$ are masses of the bottom quark, top quark and $W$
boson, and the width of the $W$ boson, respectively. 
$G_F$ and $V_{tb}$ are the Fermi constant and the Cabibbo-Kobayashi-Maskawa (CKM) matrix 
element.   
The three kinematic variables are $x=2E_l/m_t$, $y=M^2_{l\nu}/m_t^2$, and
the cosine between directions of the charged lepton and the spin axis of the
top quark. $S=1(0)$ corresponds to top quark being fully (un)polarized.  
The Dalitz variables $x$ and $y$ fulfill kinematic constraints
\begin{equation}
0\leq y \leq (1-\epsilon)^2,\,\, \omega_-\leq x\leq \omega_+,
\end{equation}
with $\omega_-=1-p_0-p_3$, $\omega_+=1-p_0+p_3$, and $p_0=(1-y+\epsilon^2)/2$,
$p_3=\sqrt{p_0^2-\epsilon^2}$. 
Distribution of transverse momentum of the charged lepton ($p_{T,l}$) defined as
with respect to axis $z$ can be derived from Eq.~(\ref{lodecay}).
For simplicity we first assume a zero width of the $W$ boson and
a massless bottom quark.
For unpolarized top quark average transverse momentum of the charged
lepton can be calculated as
\begin{equation}\label{ptave}
\langle p_{T,l}\rangle \equiv \frac{\int p_{T,l}d\Gamma}{\int d\Gamma}
={\pi\over 16}\frac{1+2\bar y+3\bar y^2}{1+2\bar y}m_t.
\end{equation}
It is straightforward to show that Eq.~(\ref{ptave}) also holds for
a polarized top quark with the spin axis not necessarily coincident
with the $z$ axis.  
Substituting mass values of $m_t=172.5$ GeV and $m_W=80.385$
GeV~\cite{Tanabashi:2018oca}, we arrive at
\begin{equation}\label{ptn1}
	\langle p_{T,l}\rangle = 37.21\,(1+0.695\frac{\delta m_t}{172.5\,{\rm GeV}})\,{\rm GeV}, 
\end{equation}   
assuming a small shift of the top-quark mass of $\delta m_t$.
Thus a 1 GeV shift of the top-quark mass translates into a
0.4\% change of the average transverse momentum.
Effects of the finite $W$ boson width can be included by integrating fully in
$y$ instead of using narrow width approximation.
The average transverse momentum is increased by one permille for a $W$ boson width of $2.2$ GeV
as compared to NWA,
\begin{equation}
\langle p_{T,l}\rangle = 37.21\,(1+0.0009\frac{\Gamma_W}{2.2\,{\rm GeV}})\,{\rm GeV}. 
\end{equation}
Effects due to the finite bottom quark mass are
expected to be even smaller with the result given by
\begin{equation}
\langle p_{T,l}\rangle = 37.21\,(1-0.0004\frac{m_b}{4.5\,{\rm GeV}})\,{\rm GeV}. 
\end{equation}

At hadron colliders kinematics of the charged lepton can be
measured in both the single top-quark production and top-quark pair
production in which the top quarks are boosted in general.
For a boost along the $z$ axis it will not affect the transverse
momentum distributions of the decay products.
Now considering the top quark travels perpendicularly to the $z$
axis with a velocity $\beta$, the average $p_{T,l}$ from the decay of an unpolarized
top quark is given by
\begin{equation}\label{ptbeta}
\langle p_{T,l}\rangle = 37.21\,{(1-0.0015\,\beta+0.257\,\beta^2)\over
\sqrt{1-\beta^2}}\,{\rm GeV}, 
\end{equation}      
as derived from Eq.~(\ref{lodecay}) keeping up to $\mathcal {O}(\beta^2)$
terms in the numerator.
At the LHC 13 TeV the top quark in $t$-channel singly production
has an average $p_T$ of about 40 GeV, while the average is about 120 GeV
in pair production.
They correspond to roughly a velocity of top quark of 0.2 and 0.6 respectively.
From direct calculations of production with subsequent leptonic
decay of the top quark at leading order (LO), we obtain the following results
for LHC 13 TeV,
\begin{equation}
\langle p_{T,l}\rangle_{t-ch} = 38.38\,{\rm GeV},\,\,
\langle p_{T,l}\rangle_{t\bar t} = 56.37\,{\rm GeV}, 
\end{equation}
which are in agreement with estimations using Eq.~(\ref{ptbeta}) together with the 
corresponding velocities at the LHC.     

From above discussion we understood the precise distributions of
the charged lepton will depend on modeling of the
top-quark kinematics and polarizations in the production.
They can be sensitive to QCD corrections in production
and in decay of the top quark.
Besides, in experimental measurements various selection
cuts are applied that can further
change distributions of the charged lepton.

\section{Theory predictions}
\label{sec:the}
In this section we present our main results on predictions of the leptonic
observables.
We first identify the signal regions used for the LHC measurement and show
the sensitivity of the proposed observable to the top-quark mass.
Then we present our theory predictions based on a next-to-next-to-leading
order calculation including decay of the top quark.
Discussions on scale variations and parametric uncertainties are also included.

\subsection{Signal regions}

In experimental measurements various kinematic cuts are applied due to
finite coverage of detectors, such as transverse momentums
or rapidities of the reconstructed jets, electrons, muons and photons.
For the single top-quark production, additional cuts or selections are
required in order to suppress SM backgrounds from the top-quark pair production
and associated production of $W/Z$ bosons and jets.
We follow closely fiducial regions used in the CMS
analyses at 8 and 13 TeV~\cite{1703.02530,1907.08330}.

We require one charged lepton in the final state with $p_T>$~26~GeV
and $|\eta|<$~2.4, and include only one family of leptons from decay of the top quark
in results through the paper unless otherwise specified.
We use the anti-$k_T$ jet algorithm~\cite{0802.1189} with a distance parameter of ${\rm D}=0.4$.
Jets are required to have $p_T>$~40 GeV and $|\eta|<$~4.7.
A clustered jet at parton level is defined as $b$-tagged if it has a non-zero
net bottom-quark number in the constituents and further has $|\eta|<$~2.4.
In addition a constant $b$-tagging efficiency of 50\% has been applied.
Light jets are defined as jets that are not $b$-tagged.
We consider two signal regions for $t$-channel production, CMS-SA and CMS-SB.
Both are required to have exactly two jets in the final state with one
being a $b$-tagged jet and the other being a light jet.
We require the transverse mass of the charged lepton and the
missing transverse momentum from neutrinos to be greater than 50~GeV.
In CMS-SB the light jet is required to stay in the forward region,
namely $|y|>$~2.5, which can further increase the signal to background
ratio.

\begin{figure}[ht]
\centering
  \includegraphics[width=0.47\textwidth]{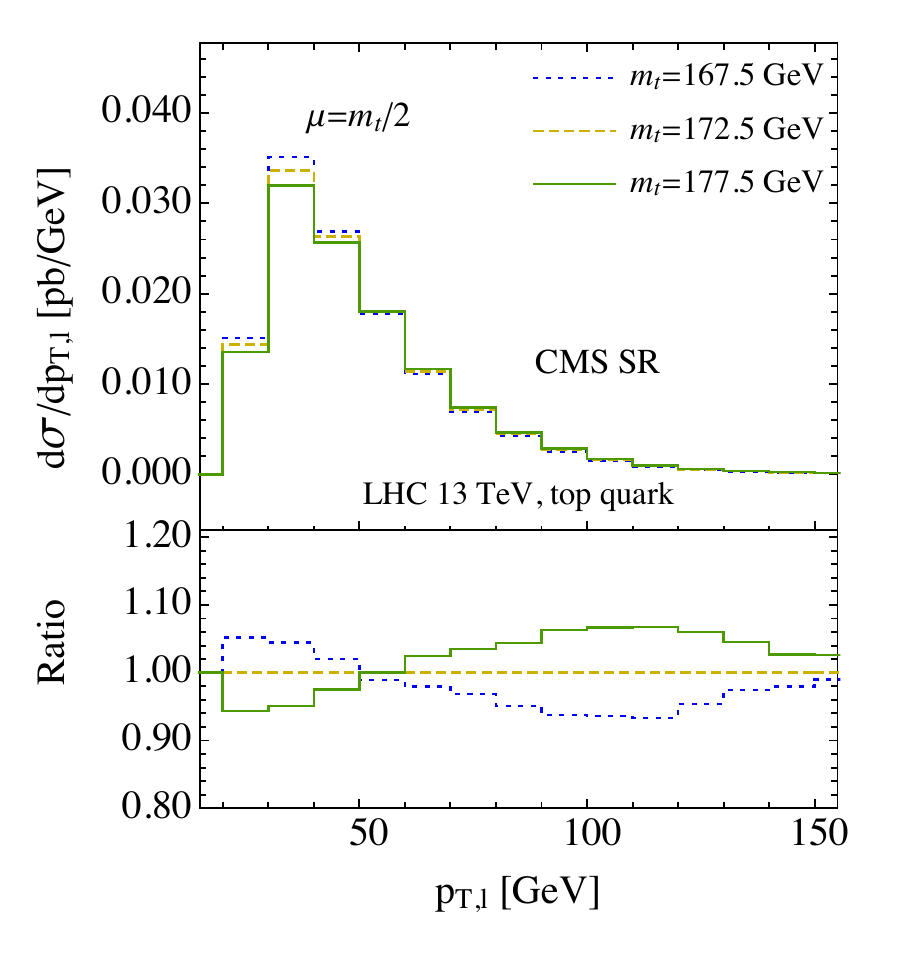}
  \hspace{0.1in}
  \includegraphics[width=0.47\textwidth]{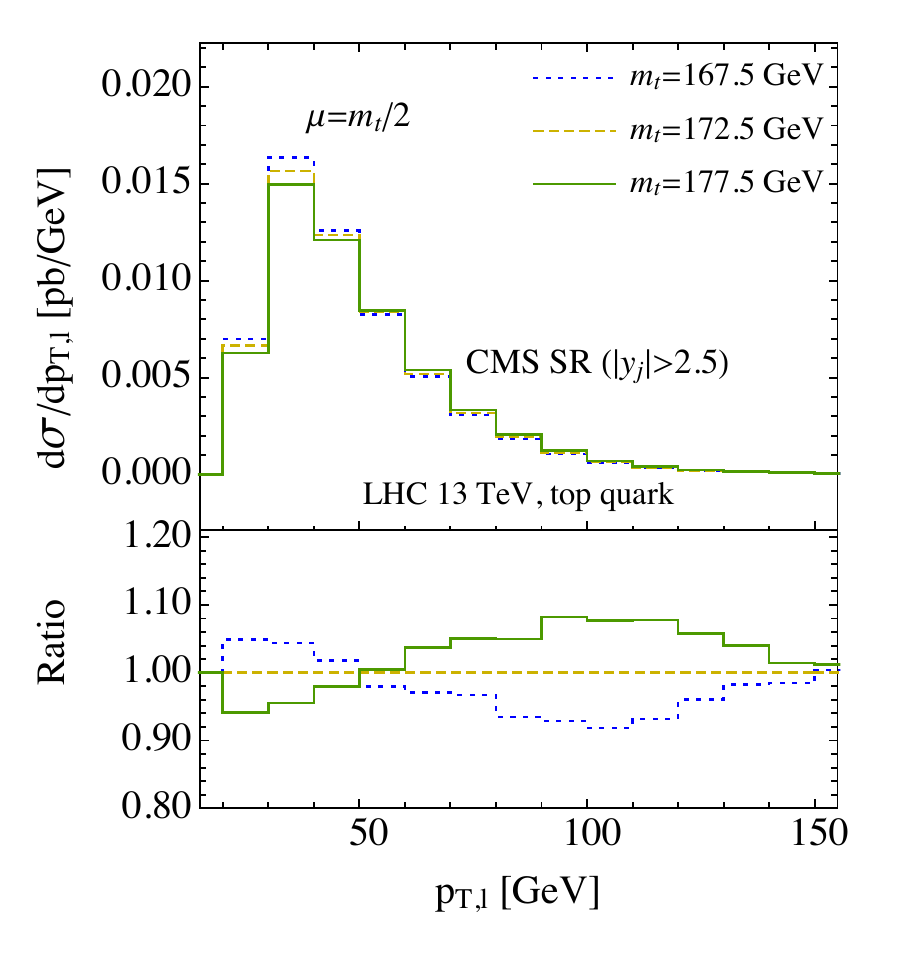}
\caption{
Transverse momentum distribution of the charged lepton within the two fiducial
regions at NLO in QCD for different choices of top-quark mass, calculated in
the 5FS for LHC 13 TeV.
\label{fig:dism}}
\end{figure}

We demonstrate the sensitivity of the leptonic distributions to
the top-quark mass in Fig.~\ref{fig:dism} for LHC 13 TeV.
We show transverse momentum distributions of the charged lepton
in the two signal regions with a top-quark mass of 172.5~GeV or
shifted by 5~GeV, calculated at NLO in QCD.
The lower inset shows ratios of the distributions with different top-quark masses.
Details of the calculation will be explained later.
The increase of top-quark mass leads to a harder $p_T$ spectrum in general.
At very large $p_T$, enhancements of the distribution
are cancelled out because of the increasing importance of
top-quark kinematics from production.
Results of two signal regions show a very similar dependence on the
top-quark mass.

\begin{figure}[ht]
\centering
  \includegraphics[width=0.6\textwidth]{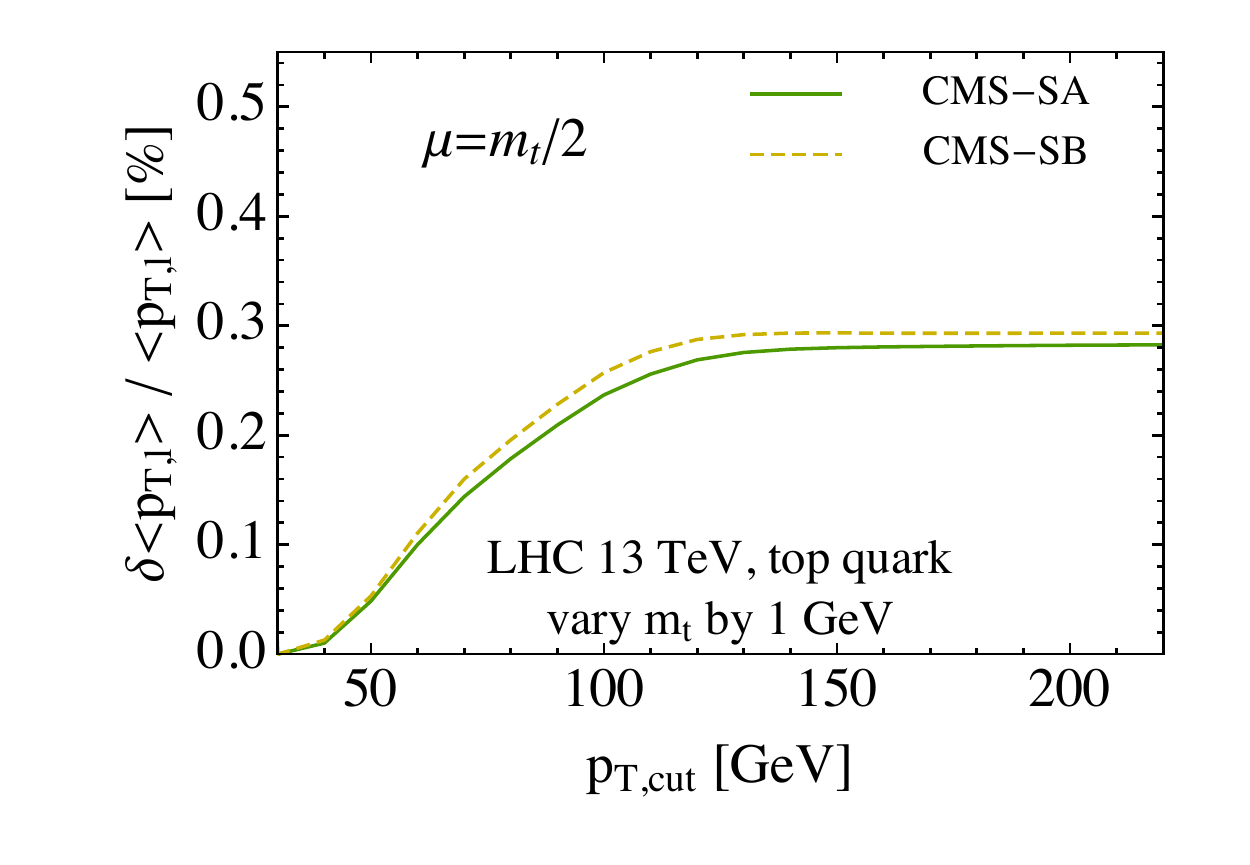}
\caption{
Induced change of the average transverse momentum of the charged lepton
within the two fiducial regions when varying the top-quark mass by 1 GeV,
as a function of an upper limit on the transverse momentum,
calculated at NLO in QCD in the 5FS for LHC 13 TeV.
\label{fig:senm}}
\end{figure}

We prefer to use a single variable to extract the top-quark mass,
rather than from a template fit to the full leptonic distribution.
We choose the variable as average $p_T$ of the charged lepton.
We can select different windows of the $p_T$ spectrum to be included.
We plot relative change of the average $p_T$ when varying the top-quark mass
by 1~GeV, as a function of an upper limit placed on $p_T$ in Fig.~\ref{fig:senm}.
For both signal regions the sensitivity saturates to a value of about 0.3\%
when the upper limit reaches above 100~GeV.
In latter sections we will present results for two representative
values of the upper limit, 100~GeV and 200~GeV, respectively.
Inclusion of high $p_T$ region usually leads to larger theory
uncertainties.

\subsection{NNLO predictions}
NNLO predictions for the $t$-channel single top-quark production in the 5-flavor
scheme are calculated using the phase-space slicing with the $N$-jettiness
variable~\cite{Stewart:2010tn,Boughezal:2015dva, Gaunt:2015pea,Berger:2016inr}
together with the method of ``projection-to-Born'' in Ref.~\cite{1506.02660}.
Details for the NNLO calculation in the 5FS can be found in Refs.~\cite{Berger:2016oht,1708.09405}.
In the calculation, QCD corrections can be
factored as from either fermion line with heavy quarks or light quarks
neglecting certain color suppressed contributions~\cite{Lindfors:1985zz,Han:1992hr,Stelzer:1997ns}.
We also include consistently the NNLO corrections in decay of the top
quark as originally calculated in~\cite{Gao:2012ja} using narrow
width approximation.
We focus on predictions for the top-quark production at LHC 13 TeV. 
Results for top anti-quark production can be obtained through
a CP transformation with substitutions of the parton distributions.

We use a PDF set of PDF4LHC15\_nnlo\_30 with
$\alpha_S(m_Z)=$~0.118~\cite{Butterworth:2015oua,Gao:2013bia,Harland-Lang:2014zoa,
Ball:2014uwa,Dulat:2015mca,Carrazza:2015aoa}, and
a nominal value of the top-quark mass of 172.5~GeV.
The central scales of QCD renormalization and factorization are
set to half of the top-quark mass.
A lower value of the QCD scale in 5FS was suggested
in Ref.~\cite{1203.6393} which shows those quasi-collinear logarithms
to be resummed are accompanied by a universal suppression from phase
space integration, as also supported by numerical calculations in~\cite{2005.12936}.
We evaluate scale uncertainty by varying the two scales independently
with a factor of two and taking the envelope of results with 9 scale
choices.
Effects due to finite width of the $W$ boson, finite mass of the
bottom quark in top-quark decay, and finite width of the top quark,
are included by adding their corrections calculated at leading order.
For example, the off-shell effects of top quark are modeled with a
Breit-Wigner shape at LO.
The resulted average $p_T$ of the charged lepton differs with that in
the NWA by 0.03 GeV, which are added into our final NNLO predictions
calculated with NWA.
We will discuss off-shell effects beyond leading order in Sec.~\ref{new}.

\begin{figure}[ht]
\centering
  \includegraphics[width=0.47\textwidth]{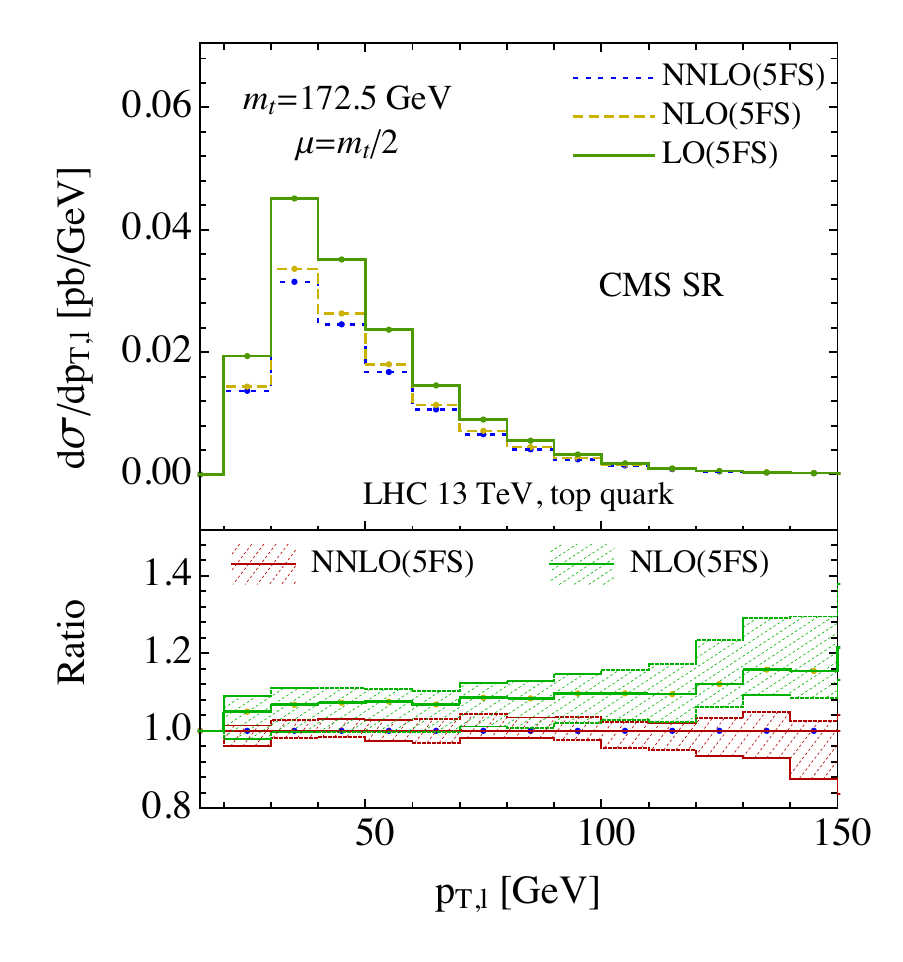}
  \hspace{0.1in}
  \includegraphics[width=0.47\textwidth]{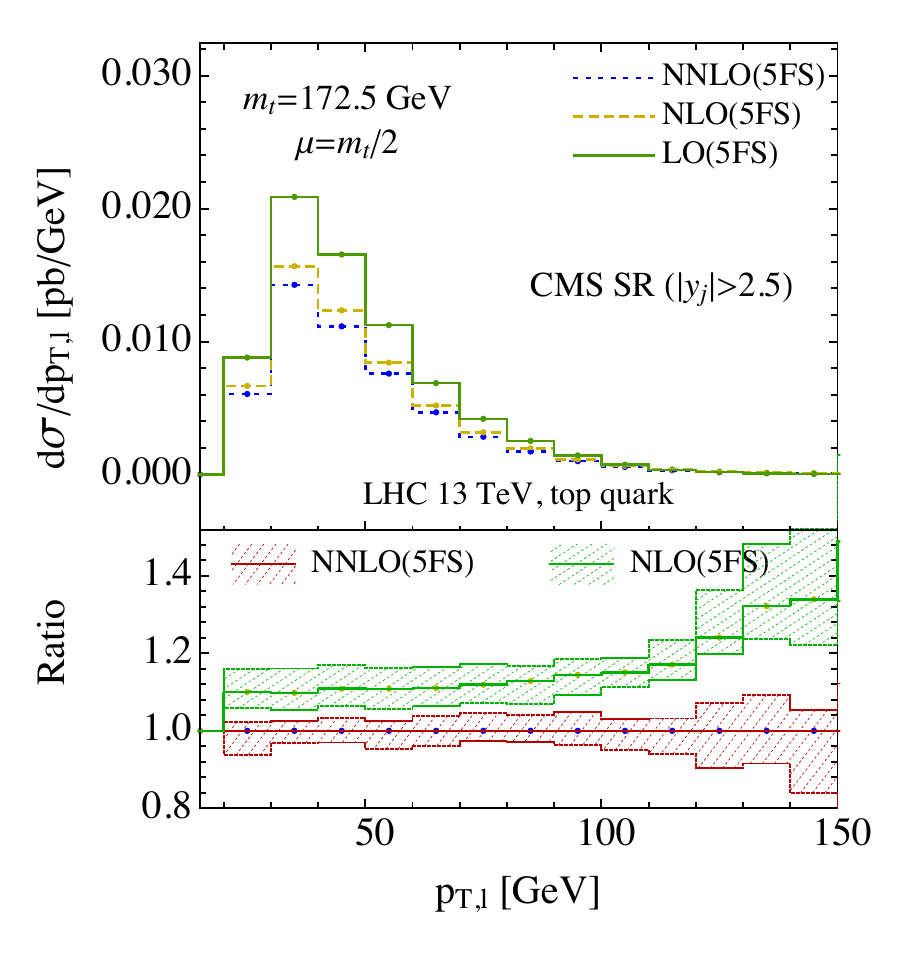}
\caption{
Transverse momentum distribution of the charged lepton within the two fiducial
regions at various orders in QCD
in the 5FS for LHC 13 TeV. Scale
variations are evaluated by taking envelop of results with 9 scale choices.
\label{fig:5f}}
\end{figure}

We show transverse momentum distributions of the charged lepton
in the two signal regions in Fig.~\ref{fig:5f} at various orders in
QCD together with scale uncertainties for LHC 13 TeV.
In the lower inset of each plot we show ratios of the predictions to
a common reference calculated at NNLO with nominal scale choice.
The NNLO corrections lead to a softer spectrum due to both soften of
the top-quark $p_T$ and additional radiations in top-quark decay~\cite{Gao:2012ja}.
Size of the NNLO corrections ranges from -5\% to -35\% for the
$p_T$ region shown.
Moderate reduction of scale uncertainties are seen when including the NNLO
corrections.
However, the scale variations at NLO underestimate the size of NNLO
corrections for the signal region CMS-SB especially in the high-$p_T$ tail.  

\begin{table}[h!]
\centering
\begin{tabular}{l|cc|cc}
\hline
	$\langle p_{T,l}\rangle $ &\multicolumn{2}{c}{CMS-SA} & \multicolumn{2}{c}{CMS-SB}\tabularnewline
\cline{2-5}
	[GeV] & $<$ 100 GeV & $<$ 200 GeV & $<$ 100 GeV & $<$ 200 GeV \tabularnewline
\hline
\hline
	LO   & 47.33$_{-0.03}^{+0.03}$(47.33) & 49.31$_{-0.08}^{+0.09}$(49.31) & 47.38$_{-0.02}^{+0.01}$(47.38) &  48.73$_{-0.04}^{+0.05}$(48.73)  \tabularnewline
\hline
	NLO  & 47.78$_{-0.14}^{+0.17}$(48.06) & 50.37$_{-0.30}^{+0.38}$(50.67) & 47.49$_{-0.09}^{+0.13}$(47.84) &  49.66$_{-0.27}^{+0.36}$(50.02)  \tabularnewline
\hline
	NNLO & 47.65$_{-0.03}^{+0.09}$(48.01) & 50.10$_{-0.16}^{+0.09}$(50.49) & 47.35$_{-0.03}^{+0.14}$(47.75) &  49.25$_{-0.12}^{+0.17}$(49.67)  \tabularnewline
\hline
\end{tabular}

\caption{
Average transverse momentum of the charged lepton within the two fiducial
regions at various orders in QCD in the 5FS for LHC 13 TeV. Scale
variations are evaluated by taking envelop of results with 9 scale choices.
Numbers in parenthesis correspond to predictions without including QCD
corrections in decay of top quark.
\label{tab:5fpt}}
\end{table}

We present detailed results on the average $p_T$ of the charged lepton in
Table~\ref{tab:5fpt} for the two signal regions and with two choices
of the upper limit on $p_T$.
Numbers in parenthesis correspond to predictions when excluding QCD corrections
in decay of the top quark, i.e., only including corrections in production of the top
quark.
We find the leading order predictions show a rather small scale variation,
which can be understood since the change of scales at LO only impact the
overall normalization and longitudinal boost of the system, not the
shape of the transverse momentum distribution.
The LO predictions can not describe well distributions of transverse momentum of
the top quark, especially at high $p_T$, as explained in~\cite{2005.12936}. 
The NNLO corrections lead to a reduction of the average $p_T$ by less than 0.2~GeV
if the upper limit of 100~GeV is applied.
The corrections are slightly larger if instead the upper limit of 200~GeV
is used.
The final NNLO predictions show scale uncertainties at the level of 0.1~GeV
which are comparable to the change as induced by a shift of the
top-quark mass of 1~GeV.
QCD corrections from top-quark decay are large comparing to our target
precision of the average $p_T$.
They reduce the average $p_T$ by about 0.3$\sim$0.4~GeV at NLO.
The NNLO corrections from top-quark decay further decrease the
average $p_T$ by 0.1~GeV in the case of signal region CMS-SA.
We recall that NNLO corrections due to top-quark decay consist
of two parts, one from pure two-loop corrections in top-quark decay and
the other from one-loop corrections in decay combined with one-loop
corrections in production.
Both of the two pieces are important. 
Cancellation between them may occur depending on the observables and
kinematic region considered.
We also show predictions on total fiducial cross sections in Table~\ref{tab:5fxsec}.
The NNLO corrections reduce the cross sections by about 6\% for
signal region CMS-SA, with predictions located at the lower boundary of
the NLO scale variations.
The reduction is about 10\% for the signal region CMS-SB.

\begin{table}[h!]
\centering
\begin{tabular}{l|cc|cc}
\hline
	$\sigma_{fid.} $ &\multicolumn{2}{c}{CMS-SA} & \multicolumn{2}{c}{CMS-SB}\tabularnewline
\cline{2-5}
	[pb] & $<$ 100 GeV & $<$ 200 GeV & $<$ 100 GeV & $<$ 200 GeV \tabularnewline
\hline
\hline
	LO   & 1.55$_{-0.21}^{+0.16}$(1.55) & 1.59$_{-0.22}^{+0.17}$(1.59) & 0.724$_{-0.08}^{+0.06}$(0.724) & 0.739$_{-0.08}^{+0.06}$(0.739)\tabularnewline
\hline
	NLO  & 1.17$_{-0.08}^{+0.03}$(1.31) & 1.22$_{-0.08}^{+0.03}$(1.36) & 0.545$_{-0.02}^{+0.02}$(0.613) & 0.562$_{-0.02}^{+0.02}$(0.632)\tabularnewline
\hline
	NNLO & 1.10$_{-0.02}^{+0.02}$(1.24) & 1.14$_{-0.02}^{+0.02}$(1.29) & 0.493$_{-0.01}^{+0.01}$(0.563) & 0.506$_{-0.01}^{+0.01}$(0.579)\tabularnewline
\hline
\end{tabular}

\caption{
Fiducial cross section at various orders in QCD in the 5FS for LHC 13 TeV. Scale
variations are evaluated by taking the envelop of results with 9 scale choices.
Numbers in parenthesis correspond to predictions without including QCD
corrections in decay of top quark.
\label{tab:5fxsec}}
\end{table}

\subsection{Parametric uncertainties}

We investigate dependence of our predictions on various inputs and
the associated parametric uncertainties.
That includes the parton distribution functions, QCD coupling constant,
and bottom quark mass.
Uncertainties due to parton distribution functions are estimated following
the PDF4LHC recommendation~\cite{Butterworth:2015oua} and using PDF4LHC15\_nnlo\_30
PDF set.
We calculate the dependence on QCD coupling constant by varying $\alpha_S(m_Z)$
by $\pm 0.0015$ from its nominal value of 0.118, and using PDFs of the same 
$\alpha_S(m_Z)$ values.
We use MMHT2014 PDF set~\cite{Harland-Lang:2015qea} with different bottom-quark masses
to calculate the changes when varying the pole mass of bottom quark by 0.5 GeV.
The impact on average $p_T$ of the charged lepton and on the total fiducial
cross section are summarized in Table~\ref{tab:par}.
We also include corresponding numbers when varying the mass of top quark by
1 GeV for comparison.

We find in all cases the parametric uncertainties on average $p_T$ are at the level of 0.01$\sim$0.02 GeV,
and are small comparing to the dependence on the top-quark mass.
On another hand, the total fiducial cross sections show larger uncertainties.
For example, the PDF uncertainties are 2\%$\sim$4\%, and the uncertainties due to
bottom quark mass are 1\%$\sim$3\% if taking error of bottom quark mass as 0.2 GeV~\cite{Harland-Lang:2015qea}.
The total fiducial cross sections are insensitive to the
top-quark mass, unlike the average $p_T$ of the charged lepton.

\begin{table}[h!]
\centering
\begin{tabular}{l|c|cc|cc}
\hline
	[GeV]/[pb]&&\multicolumn{2}{c}{CMS-SA} & \multicolumn{2}{c}{CMS-SB}\tabularnewline
\cline{3-6}
	&& $<$ 100 GeV & $<$ 200 GeV & $<$ 100 GeV & $<$ 200 GeV \tabularnewline
\hline
\hline
	\multirow{2}{*}{PDFs(68\% C.L.)} &$\delta\langle p_{T,l}\rangle $  & 0.014 & 0.023 & 0.021 & 0.022 \tabularnewline
	& $\delta\sigma_{fid}$  & 0.040 & 0.041 & 0.020 & 0.021 \tabularnewline
\hline
	\multirow{2}{*}{$\alpha_S(m_Z)$(0.0015)} &$\delta\langle p_{T,l}\rangle $  & $<$ 0.01 & $<$ 0.01 & $<$ 0.01 & $<$ 0.01 \tabularnewline
	& $\delta\sigma_{fid}$   & 0.017 & 0.018 & 0.005 & 0.005\tabularnewline
\hline
	\multirow{2}{*}{$m_b$(0.5 GeV)} &$\delta\langle p_{T,l}\rangle $  & $<$ 0.01 & $<$ 0.01 & $<$ 0.01 & $<$ 0.01 \tabularnewline
	& $\delta\sigma_{fid}$   & 0.064 & 0.066 & 0.029 & 0.030\tabularnewline
\hline
	\multirow{2}{*}{$m_t$(1.0 GeV)} &$\delta\langle p_{T,l}\rangle $  & 0.11 & 0.14 & 0.12 & 0.14 \tabularnewline
	& $\delta\sigma_{fid}$   & 0.0039 & 0.0035 & 0.0013 & 0.0011\tabularnewline
\hline
\end{tabular}

\caption{
PDF uncertainties on the average transverse momentum of the charged lepton
and on the fiducial cross section within the two fiducial regions, followed 
by induced changes on the same quantities when varying $\alpha_S(m_Z)$, $m_b$,
and $m_t$ by the amount in parenthesis. 
\label{tab:par}}
\end{table}

\section{Alternative theories}
\label{sec:alt}

We present two alternative theory predictions concerning both the perturbative
and non-perturbative components in $t$-channel production of single top quark.
Comparison with our nominal predictions can lead a better understanding
on the related theoretical uncertainties.

\subsection{Heavy-quark schemes}
It is known that the $t$-channel production can also be calculated in
a factorization scheme with a fixed 4 light-quark flavors.
The 5FS has the advantages of resumming
large logarithms of bottom-quark mass due to gluon splitting
into bottom quarks from the initial state.
The 4FS maintains full bottom quark mass dependence through fixed order
with current predictions available only at NLO in QCD.
We note that leading order calculations in 4FS already
contain ingredients appearing at next-to-leading order in 5FS.

Critical questions arise on the use and agreement of the two
heavy-quark schemes in $t$-channel single top-quark production, with 
efforts at understanding made in Refs.~\cite{0903.0005,1203.6393,1711.02568}.
In a recent study by one of the authors~\cite{2005.12936}, we compare predictions at
NNLO in 5FS to those at NLO in 4FS without decaying of the top quark.
We found the two schemes agree within a few percent in general
for the shape of kinematic distributions of the top quark, and differ
on the overall normalizations.
We conclude that 5FS provides a better modeling on $t$-channel
production when both are evaluated at comparable perturbative orders.
Here we extend the comparison to include leptonic decay of
the top quark, focusing on the leptonic observables discussed.

We use MCFM~\cite{Campbell:2016jau,Boughezal:2016wmq} program to calculate $t$-channel
single top-quark production with subsequent decays in the 4FS.
The original calculation was detailed in Ref.~\cite{0903.0005}.
We use CT14 NNLO PDFs~\cite{Dulat:2015mca} with 4 light-quark flavors through
the comparison and a bottom-quark mass of 4.75~GeV.
We set the nominal QCD renormalization scale and factorization scale to 
half of the top-quark mass.
Scale variations are evaluated with the 9-scales envelope same as before.

\begin{figure}[ht]
\centering
  \includegraphics[width=0.47\textwidth]{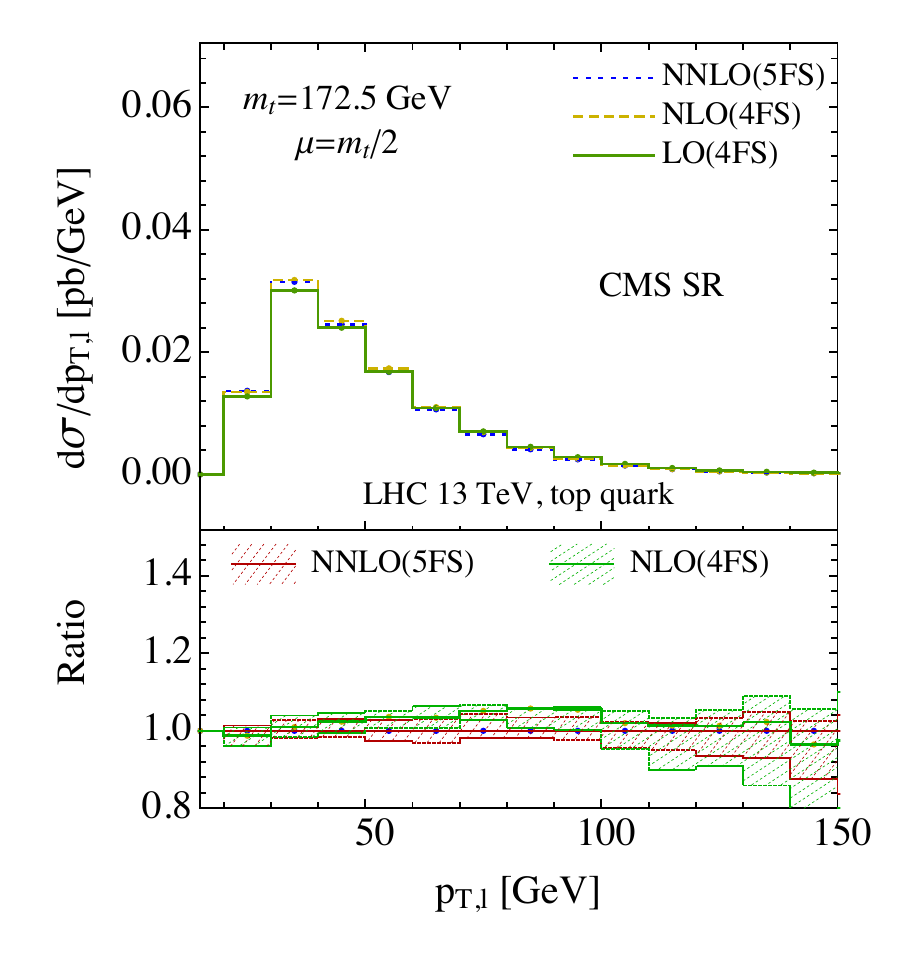}
  \hspace{0.1in}
  \includegraphics[width=0.47\textwidth]{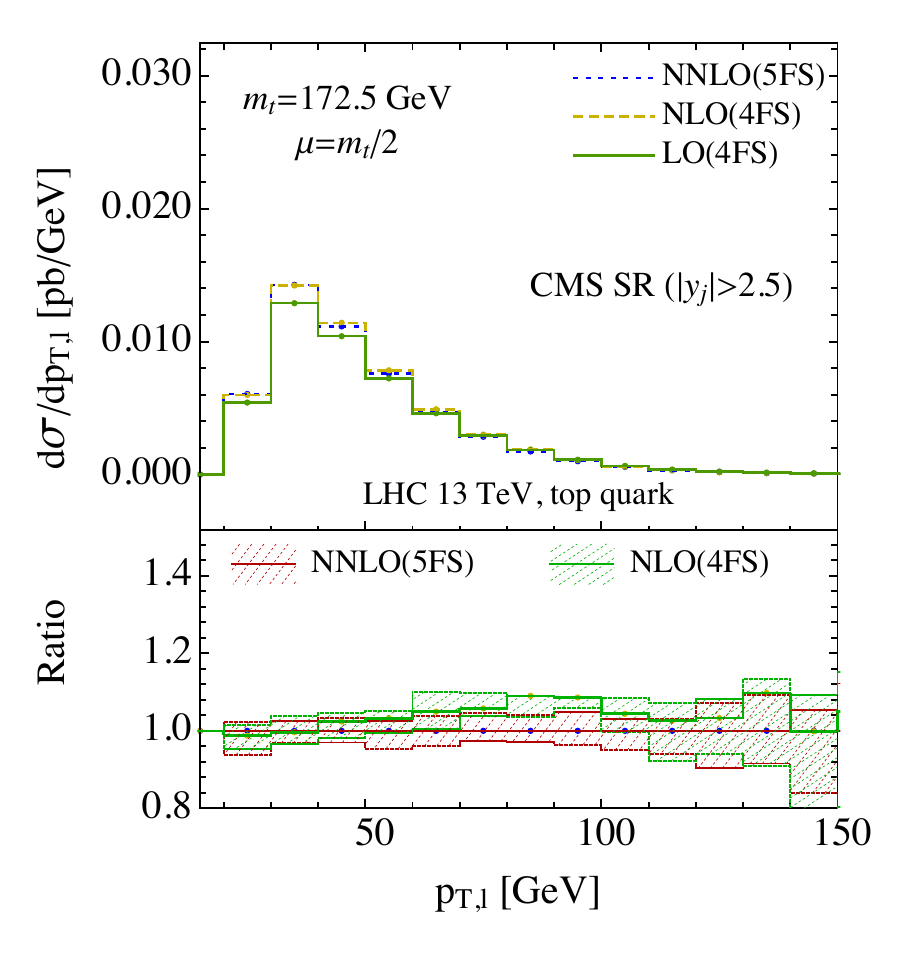}
\caption{
Transverse momentum distribution of the charged lepton within the two fiducial
regions at various orders in QCD for LHC 13 TeV,
comparing 5FS with 4FS. Scale variations are evaluated by taking the envelop of
results with 9 scale choices.
\label{fig:fo4f}}
\end{figure}

We show 4FS predictions on transverse momentum distributions of the charged lepton
in Fig.~\ref{fig:fo4f} for the two fiducial regions, compared with
the NNLO predictions in 5FS.
In the comparison the NLO predictions in 4FS include the NLO corrections
in top-quark decay, and the NNLO predictions in 5FS include further NNLO
corrections in decay.
We find the LO and NLO predictions in 4FS show less differences as
compared to the case of stable top quark in Ref.~\cite{2005.12936}.
That is because of the jet veto condition applied, namely
requiring exactly two jets in the final state.
For the same reason the NLO predictions in 4FS agree well 
with NNLO ones in 5FS even for the overall normalizations.
Shape differences of the two predictions can be understood as due
to both the harder $p_T$ spectrum of the top quark from production
in 4FS and the inclusion of NNLO corrections from decay
in 5FS.
The scale variations are slightly larger in the high $p_T$ region
for the 4FS predictions.

\begin{table}[h!]
\centering
\begin{tabular}{l|cc|cc}
\hline
	$\langle p_{T,l}\rangle $ &\multicolumn{2}{c}{CMS-SA} & \multicolumn{2}{c}{CMS-SB}\tabularnewline
\cline{2-5}
	[GeV] & $<$ 100 GeV & $<$ 200 GeV & $<$ 100 GeV & $<$ 200 GeV \tabularnewline
\hline
\hline
	LO(4FS)   & 48.45$_{-0.05}^{+0.02}$(48.45) & 51.65$_{-0.14}^{+0.12}$(51.65) & 48.22$_{-0.06}^{+0   }$(48.22) &  50.79$_{-0.10}^{+0   }$(50.79)  \tabularnewline
\hline
	NLO(4FS)  & 47.94$_{-0.18}^{+0.08}$(48.11) & 50.33$_{-0.49}^{+0.23}$(50.55) & 47.79$_{-0.20}^{+0.07}$(47.89) &  49.72$_{-0.44}^{+0.20}$(49.83)  \tabularnewline
\hline
	NNLO(5FS) & 47.65$_{-0.03}^{+0.09}$(48.01) & 50.10$_{-0.16}^{+0.09}$(50.49) & 47.35$_{-0.03}^{+0.14}$(47.75) &  49.25$_{-0.12}^{+0.17}$(49.67)  \tabularnewline
\hline
\end{tabular}

\caption{
Average transverse momentum of the charged lepton within the two fiducial
regions at various orders in QCD for LHC 13 TeV, comparing 5FS with 4FS. Scale
variations are evaluated by taking the envelop of results with 9 scale choices.
Numbers in parenthesis correspond to predictions without including QCD
corrections in decay of top quark.
\label{tab:4fpt}}
\end{table}

More comparisons can be found in Table~\ref{tab:4fpt} for
the average $p_T$ of the charged lepton.
The numbers in parenthesis correspond to predictions without including QCD
corrections in decay of top quark.
We find scale variations of LO predictions in 4FS are small
and underestimate the genuine NLO corrections especially when including
high-$p_T$ regions, similar to the case of LO predictions in 5FS.
The difference on average $p_T$ between NLO predictions in 4FS and NNLO
predictions in 5FS is about 0.3 GeV for the fiducial region CMS-SA
and 0.4 GeV for the fiducial region CMS-SB.
Half of the difference can be attributed to the different treatment
on corrections in decay of the top quark.
Scale variations are slightly larger for average $p_T$ from NLO predictions
in 4FS.
Predictions of the two schemes overlap in general once considering both scale
variations.
Similar results for the total fiducial cross sections are shown in
Table~\ref{tab:4fxsec} where even better agreement are seen
between the two schemes.

\begin{table}[h!]
\centering
\begin{tabular}{l|cc|cc}
\hline
	$\sigma_{fid.} $ &\multicolumn{2}{c}{CMS-SA} & \multicolumn{2}{c}{CMS-SB}\tabularnewline
\cline{2-5}
	[pb] & $<$ 100 GeV & $<$ 200 GeV & $<$ 100 GeV & $<$ 200 GeV \tabularnewline
\hline
\hline
	LO(4FS)   & 1.08$_{-0.12}^{+0.16}$(1.08) & 1.13$_{-0.13}^{+0.16}$(1.13) & 0.464$_{-0.06}^{+0.07}$(0.464) & 0.480$_{-0.06}^{+0.08}$(0.480)\tabularnewline
\hline
	NLO(4FS)  & 1.12$_{-0.02}^{+0.02}$(1.18) & 1.16$_{-0.02}^{+0.02}$(1.22) & 0.503$_{-0.01}^{+0.01}$(0.528) & 0.517$_{-0.01}^{+0.01}$(0.543)\tabularnewline
\hline
	NNLO(5FS) & 1.10$_{-0.02}^{+0.02}$(1.24) & 1.14$_{-0.02}^{+0.02}$(1.29) & 0.493$_{-0.01}^{+0.01}$(0.563) & 0.506$_{-0.01}^{+0.01}$(0.579)\tabularnewline
\hline
\end{tabular}

\caption{
Fiducial cross section at various orders in QCD for LHC 13 TeV, comparing 4FS with 5FS.
Scale variations are evaluated by taking the envelop of results with 9 scale choices.
Numbers in parenthesis correspond to predictions without including QCD
corrections in decay of top quark.
\label{tab:4fxsec}}
\end{table}

\subsection{Parton shower and hadronization}
We compare our parton-level results with those from various Monte Carlo
generators both in 5-flavor number scheme.
We calculate the fiducial cross sections and distributions at NLO in QCD
matched with parton shower using MG5\_aMC$@$NLO program~\cite{Alwall:2014hca}.
We generate matched events with stable top quarks that are further decayed
with MadSpin~\cite{Artoisenet:2012st}.
Events are then passed to various generators for parton shower
and hadronization, including PYTHIA6~\cite{Sjostrand:2006za}, PYTHIA8~\cite{Sjostrand:2014zea}, and
HERWIG7~\cite{Bahr:2008pv}.
Finally the events are analysed with MadAnalysis5~\cite{Conte:2012fm} and FastJet~\cite{Cacciari:2011ma}.
We use same input parameters as in previous fixed-order calculations for
PDFs, QCD scales and selection cuts.
We have checked the total inclusive cross sections agree at NLO in the two calculations.  
In MC simulations one difference with respect to fixed-order calculation is on definition of
$b$-tagged jet for which we use the default method implemented in MadAnalysis5.
Be specific, for each event after jet clustering, one searches for intermediate $b$
quarks in the MC record.
The clustered jets are considered as a $b$-tagged jet if it can be associated with
a MC $b$ quark inside the jet cone.
Similarly a $b$-tagging efficiency of 50\% is applied on the $b$-tagged jet.
\begin{figure}[ht]
\centering
  \includegraphics[width=0.47\textwidth]{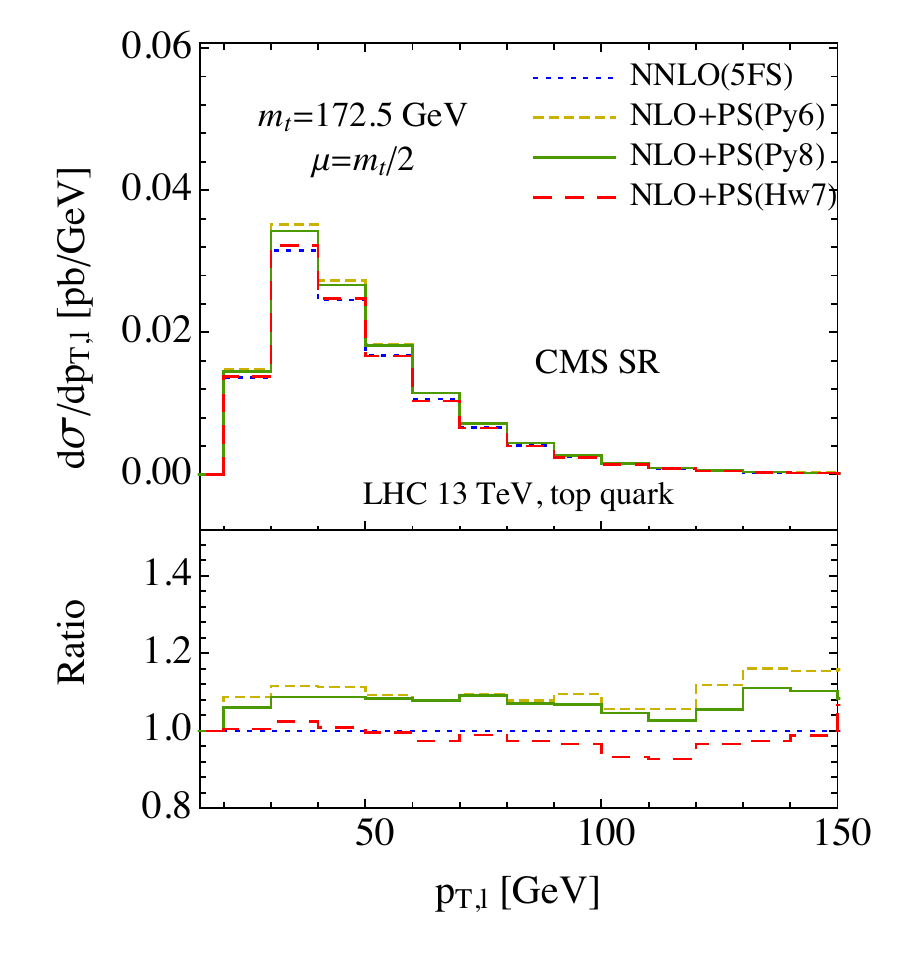}
  \hspace{0.1in}
  \includegraphics[width=0.47\textwidth]{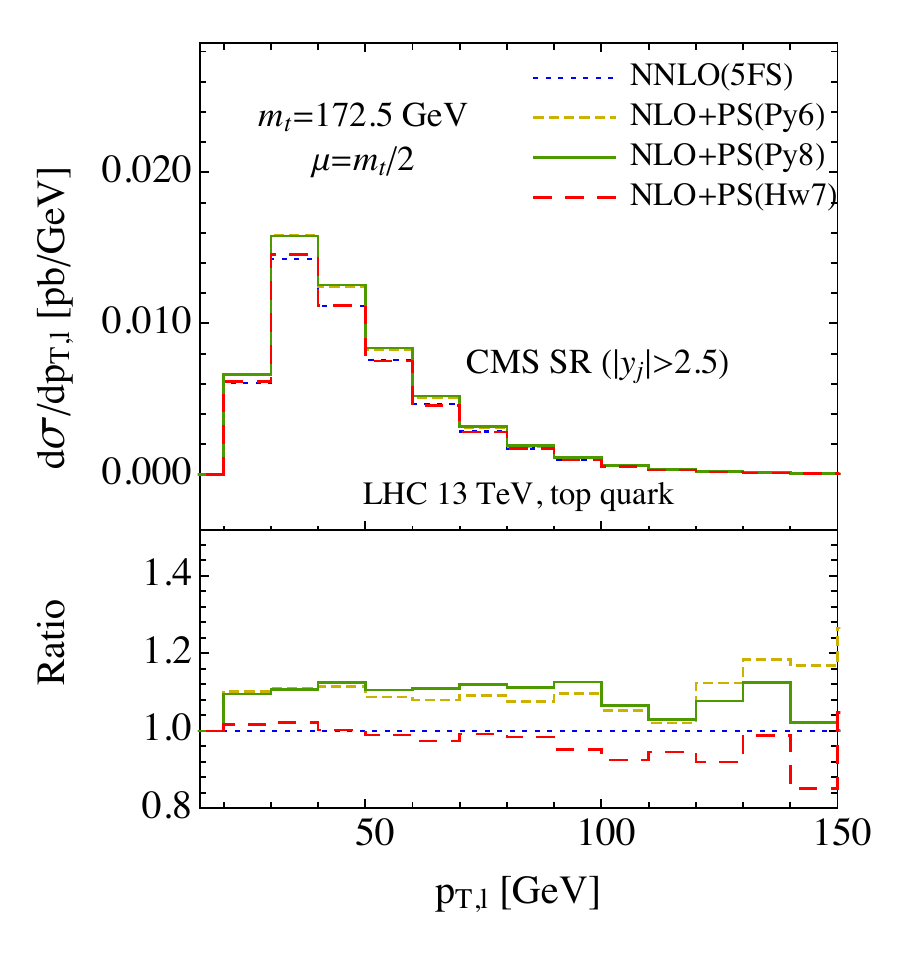}
\caption{
Transverse momentum distribution of the charged lepton within the two fiducial
regions, comparing predictions at fixed-order and those from various
event generators in the 5FS for LHC 13 TeV.
\label{fig:fops}}
\end{figure}

We show MC predictions on transverse momentum distributions of the charged lepton
in Fig.~\ref{fig:fops} for the two fiducial regions, comparing with
the NNLO predictions calculated earlier.
We find HERWIG7 provides different predictions comparing with
PYTHIA6 and PYTHIA8 while the latter two show very good agreement.
That can be due to different shower algorithms or possibly the way
of different shower programs handling decayed resonance, for example
as studied in Refs.~\cite{1603.01178,1607.04538}.
We can also compare MC predictions with the NNLO predictions.
It is interesting that the two show very good agreement on shape of
the distribution though the normalization is higher by about 10\% for
PYTHIA6 and PYTHIA8.
Such agreement is non-trivial since the matrix elements used in
MC predictions do not include NLO corrections in top-quark
decay which have large impact on the $p_T$ distribution as shown
in Table~\ref{tab:5fpt}. 
Parton shower resummation takes into account part
of the missing NLO and NNLO corrections and brings the MC predictions
closer to the NNLO fixed-order predictions.
Comparison on the average $p_T$ of the charged lepton are
summarized in Table~\ref{tab:mcpt} along with the fiducial
cross section shown in Table~\ref{tab:mcxsec}.
In both tables we also include NLO fixed-order predictions with and
without corrections in decay of the top quark.
As a cross check we show the NLO predictions calculated in 5FS with MCFM
and NWA, which agree well with our results.
The small differences are
due to off-shell effects at LO included in our results as mentioned earlier.
For MC predictions, numbers in parenthesis
correspond to turning hadronizaiton off in the generators.
The hadronization corrections are small for all generators
considered, within the statistical uncertainties which are about 0.03 GeV for
the average $p_T$.
Hadronization reduces the fiducial cross section by a few percents
for HERWIG7. 
From Table~\ref{tab:mcxsec} we find normalizations of MC predictions
generally lie between NLO predictions with and without corrections in
top-quark decay.
On the other hand, for the average $p_T$, MC predictions are closer to the
NNLO predictions.
The MC predictions on the average $p_T$ from different parton showers show
a spread of $0.2\sim 0.4$ GeV due to different approximations used for
higher-order QCD corrections.
That is not surprising since the NLO QCD corrections from top quark decay
alone can induce a shift of similar size.
We note that there exist MC generators including
full NLO QCD corrections~\cite{1603.01178,1907.12586} which will
be discussed in the following section.
In the future once a calculation at NNLO matched with parton
shower becomes available we expect the dependence on parton showers
can be largely reduced.

\begin{table}[h!]
\centering
\begin{tabular}{l|cc|cc}
\hline
	$\langle p_{T,l}\rangle $ &\multicolumn{2}{c}{CMS-SA} & \multicolumn{2}{c}{CMS-SB}\tabularnewline
\cline{2-5}
	[GeV] & $<$ 100 GeV & $<$ 200 GeV & $<$ 100 GeV & $<$ 200 GeV \tabularnewline
\hline
\hline
	PYTHIA8   & 47.66(47.69)	 & 50.05(50.06)	 & 47.43(47.47)       &  49.25(49.30)	  \tabularnewline
\hline
	PYTHIA6   & 47.54(47.51)	 & 50.01(49.97)	 & 47.23(47.24)       &  49.15(49.18)	  \tabularnewline
\hline
	HERWIG7   & 47.40(47.38)	 & 49.77(49.66)	 & 47.09(47.14)       &  48.90(48.89)	  \tabularnewline
\hline
	NNLO & 47.65 & 50.10 & 47.35 &  49.25  \tabularnewline
\hline
	NLO(w/o decay) & 47.78(48.06) & 50.37(50.67) & 47.49(47.84) &  49.66(50.02)  \tabularnewline
\hline
	MCFM(w/o decay) & 47.81(48.08) & 50.40(50.70) & 47.52(47.87) &  49.69(50.04)  \tabularnewline
\hline
\end{tabular}

\caption{
Average transverse momentum of the charged lepton within the two fiducial
regions, comparing predictions at fixed-order and those from various
event generators in the 5FS for LHC 13 TeV.
Numbers in parenthesis correspond to MC predictions without including
hadronization or fixed-order predictions without including QCD corrections
in decay of top quark. 
\label{tab:mcpt}}
\end{table}

\begin{table}[h!]
\centering
\begin{tabular}{l|cc|cc}
\hline
	$\sigma_{fid.} $ &\multicolumn{2}{c}{CMS-SA} & \multicolumn{2}{c}{CMS-SB}\tabularnewline
\cline{2-5}
	[pb] & $<$ 100 GeV & $<$ 200 GeV & $<$ 100 GeV & $<$ 200 GeV \tabularnewline
\hline
\hline
	PYTHIA8  & 1.23(1.24) & 1.27(1.28) & 0.565(0.570) & 0.580(0.580)\tabularnewline
\hline
	PYTHIA6  & 1.24(1.25) & 1.29(1.29) & 0.555(0.559) & 0.570(0.573)\tabularnewline
\hline
	HERWIG7  & 1.14(1.17) & 1.17(1.21) & 0.510(0.524) & 0.535(0.548)\tabularnewline
\hline
	NNLO     & 1.10 & 1.14 & 0.493 & 0.506\tabularnewline
\hline
	NLO(w/o decay)     & 1.17(1.31) & 1.22(1.36) & 0.545(0.613) & 0.562(0.632)\tabularnewline
\hline
\end{tabular}

\caption{
Fiducial cross sections within the two fiducial
regions, comparing predictions at fixed-order and those from various
event generators in the 5FS for LHC 13 TeV.
Numbers in parenthesis correspond to MC predictions without including
hadronization or fixed-order predictions without including QCD corrections
in decay of top quark. 
\label{tab:mcxsec}}
\end{table}

\section{Discussions}
\label{sec:dis}

In this section we further discuss several theory and experimental subjects
which are relevant for extraction of the top-quark mass.
That includes impact of various experimental selections, for example, contributions
from leptonic decay of $\tau$ lepton in top-quark decay, isolation of lepton
from jets, and $b$-tagging efficiency.
Theory topics include contributions of  non-resonant diagrams,
non-factorized corrections, and electroweak corrections.
In addition, we estimate various standard model backgrounds and
propose a possible solution on reducing their impact.
Results shown here are calculated with MG5 at leading order matched
with parton shower and hadronization via PYTHIA6 unless otherwise specified.

\subsection{Signal selection and corrections}

We start with contributions from leptonic decay of $\tau$ lepton.
They can be counted as either part of the signals or a background
to be subtracted.
The inclusive cross sections from $\tau$ decay are suppressed by a branching
ratio of 17\%.
In the fiducial regions selected, the $\tau$ contributions are further
suppressed due to the $p_T$ threshold of charged lepton as well as the cut on transverse mass,
since more neutrinos are presented in final state.
For the same reason it has a softer spectrum for the charged lepton
comparing to those from direct production.
As shown in Table~\ref{tab:cor}, the $\tau$ contributions amount to about
2\% of the direct contributions for the fiducial cross sections, and reduce
the average $p_T$ of charged lepton by 0.1~GeV.

\begin{table}[h!]
\centering
\begin{tabular}{l|c|cc|cc}
\hline
 [GeV]/[pb]	&&\multicolumn{2}{c}{CMS-SA} & \multicolumn{2}{c}{CMS-SB}\tabularnewline
\cline{3-6}
	&& $<$ 100 GeV & $<$ 200 GeV & $<$ 100 GeV & $<$ 200 GeV \tabularnewline
\hline
\hline
	\multirow{2}{*}{$\tau$ decay} &$\delta\langle p_{T,l}\rangle $  & -0.13 & -0.15 & -0.13 & -0.15 \tabularnewline
	& $\delta\sigma_{fid}$  & 0.026 & 0.026 & 0.012 & 0.012 \tabularnewline
\hline
	\multirow{2}{*}{lepton isolation} &$\delta\langle p_{T,l}\rangle $  & 0.10 & 0.10 & 0.05 & 0.06 \tabularnewline
	& $\delta\sigma_{fid}$   & -0.017 & -0.017 & -0.003 & -0.003\tabularnewline
\hline
	$b$-tagging (40\%) &$\delta\langle p_{T,l}\rangle $  & 0.08 & 0.13 & 0.005 & 0.007 \tabularnewline
\hline
	\multirow{2}{*}{non-resonant} &$\delta\langle p_{T,l}\rangle $  & 0.15 & 0.26 & 0.02 & 0.06 \tabularnewline
	& $\delta\sigma_{fid}$   & -0.018 & -0.018 & 0.008 & 0.008\tabularnewline
\hline
\end{tabular}

\caption{
Changes of the average transverse momentum of the charged lepton
and of the fiducial cross section within the two fiducial regions,
when including contributions from $\tau$ decay, applying lepton
isolation, varying $b$-tagging efficiency, and including non-resonant
contributions.
\label{tab:cor}}
\end{table}
In previous calculations we have not applied any isolation cuts on the charged
lepton from jets, which are usually imposed in experimental analyses.
We repeat our NLO calculations by further requiring $\Delta R_{lj(b)}>0.4$.
The changes on fiducial cross section and average $p_T$ are summarized 
in Table~\ref{tab:cor}.
The isolation cut has less impact when requiring the light jet in forward
region, i.e., for the signal region CMS-SB, since then the charged lepton
is unlikely to be close to the light jet.
We also vary the $b$-tagging efficiency from our nominal choice of
50\% to 40\%.
That leads to an overall rescaling of the cross section and
distributions except if there exist more than one true $b$-jets
in the final state, which is the case for beyond leading order.
By repeating our NLO calculations we found the changes on average $p_T$
is negligible for the signal region CMS-SB since it is unlikely the light jet
is due to mistagging of true $b$-jet.
In reality nonuniformity of $b$-tagging efficiency
may lead to further changes of the average $p_T$ due to correlations between
kinematics of the $b$ quark and of the charged lepton.
Next we move to various theory aspects starting with non-resonant
contributions, namely production of $W^+bj$ via electroweak interactions
without a top-quark resonance.
Those non-resonant diagrams can interfere with the resonant diagrams
and induce non-negligible contributions as shown in Table~\ref{tab:cor}.
The effects are much smaller in the fiducial region CMS-SB due to
non-forward nature of the light jet in non-resonant production.
We should mention that there are also non-resonant diagrams of $W^+bj$
production from QCD interactions.
They do not interfere with the others at LO and we leave them to the
$W^+JJ$ category that will be discussed later in the background section.
\begin{figure}[ht]
\centering
  \includegraphics[width=0.47\textwidth]{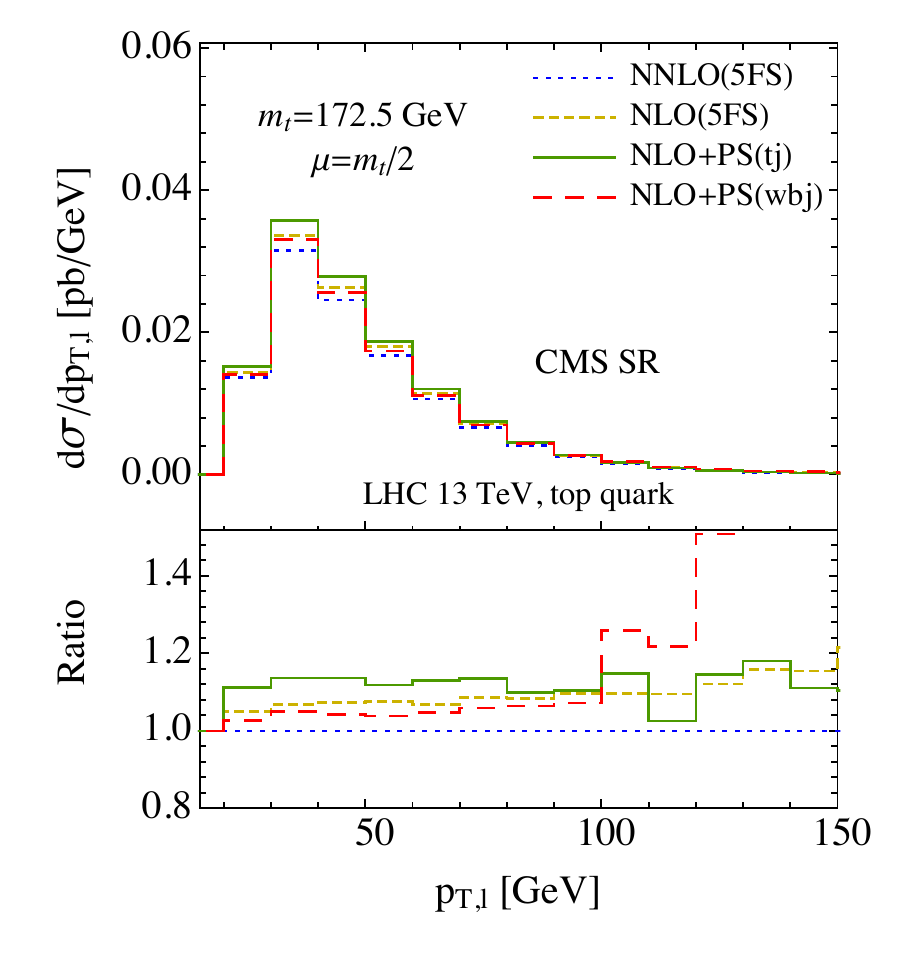}
  \hspace{0.1in}
  \includegraphics[width=0.47\textwidth]{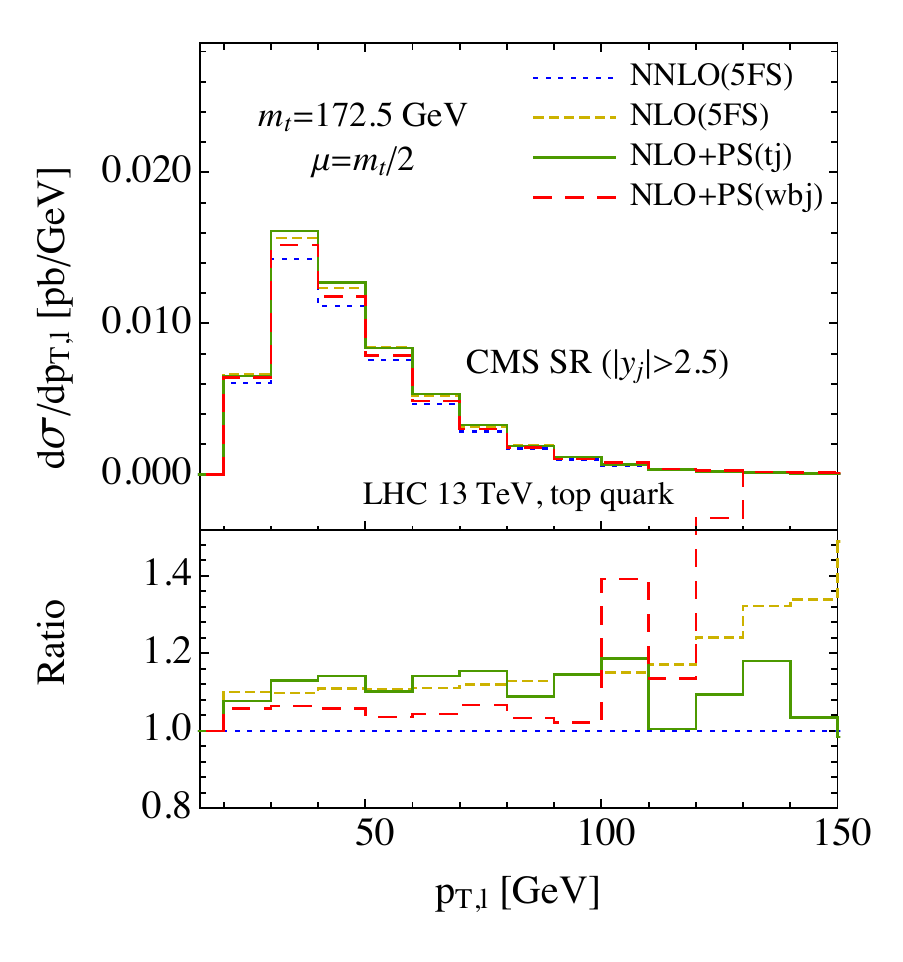}
\caption{
Transverse momentum distribution of the charged lepton within the two
fiducial regions, comparing predictions at fixed-order and those from
event generators $tj$ and $Wbj$ (see text for details) in the 5FS for LHC 13 TeV.
\label{fig:fnon}}
\end{figure}

\subsection{Non-factorized and EW corrections}\label{new}
The NNLO predictions presented are based on a calculation using NWA
and structure function approach.
There exist missing QCD corrections due to non-factorized
diagrams starting at NLO, e.g., with a gluon connecting
bottom quarks in production and in decay of the top quark.
Those non-factorized corrections have been studied in details
in~\cite{1305.7088,Neumann:2019kvk}.
They are formally of the size $\alpha_S\Gamma_t/m_t$, namely suppressed
by the width of the top quark, but can be enhanced in certain kinematic region.
We follow the strategy in Ref.~\cite{1907.12586} on identifying the corrections.
We calculate the full NLO predictions for the production of final
state $W^+bj$ and then subtract the contributions due to $s$-channel single top production
and associated production of $tW$.
All contributions are calculated using MG5\_aMC$@$NLO with parton shower and
hadronization applied from PYTHIA6, instead of calculated at fixed order.

The resulting transverse momentum distributions of the charged lepton
are presented in Fig.~\ref{fig:fnon}, comparing with fixed-order results
and MC result with PYTHIA6 shown in early sections.
The two MC predictions are denoted as $tj$ and $W^+bj$ respectively.
In comparison the latter includes exact NLO corrections in decay of
the top quark and further the non-factorized corrections just mentioned.
For transverse momentum below 100~GeV we find the $W^+bj$ predictions locate
well in between previous NLO and NNLO fixed-order predictions.
We find a trend of large enhancement of the distribution at beyond
100~GeV in $W^+bj$ predictions though accompanied with significant MC statistical
errors.
In Ref.~\cite{1907.12586} it also shows the NLO electroweak corrections can
induce a significant change on shapes of various distributions.
For comparison to experimental data, a recalculation of the
full NLO QCD and EW corrections focusing on transverse momentum of the
charged lepton will be desirable.
That can be done following various techniques outlined in Ref.~\cite{1907.12586}
and with a careful separation of backgrounds that are already accounted
for in the experimental analyses.

Finally there are also non-factorized NNLO QCD corrections in production stage which are
beyond the structure function approach, for example, from the double-box
diagrams and also interferences of $t$-channel and $s$-channel at
NNLO.
It is not clear how they may change shape of various distributions.
Giving the fact that they are suppressed by QCD colors and also considering
the size of the known NNLO corrections in production, we estimate their impact to the
average $p_T$ to be within the scale variations considered.
Similar non-factorized corrections also exist in production of the Higgs
boson and are estimated using eikonal approximation~\cite{Liu:2019tuy,Dreyer:2020urf}.

\subsection{Backgrounds}
The SM backgrounds mainly consist of top-quark pair production,
single top-quark production in $s$-channel and in associated with a $W$ boson,
QCD production of $WJJ$, and diboson production.
For QCD production of $WJJ$, the jet $J$ can arise from not only a
bottom quark, but also a charm quark, even a gluon or a light quark.
In the latter it mimics the signal due to mistagging
which we choose a rate of 3\% for charm quark and 0.1\% for gluon and light
quarks~\cite{1907.08330}.
We summarize the fiducial cross sections of various backgrounds and the average
transverse momentum in Table~\ref{tab:bk}.
For comparison we also show backgrounds for top anti-quark production, and
include numbers for signal process as well which are calculated at NNLO.
The average momentums are calculated from spectrums of individual signals
and backgrounds in the fiducial regions.
They are close in size for signal and backgrounds due to
the same kinematic selections used, for instance, a threshold of 26 GeV
on $p_T$ of the charged lepton.
In Table~\ref{tab:bk} we do not repeat the row if the background
contributes equally to top quark and anti-quark processes.

In general the top-quark pair production is dominant among all backgrounds
due to the large cross section, though it requires the additional charged lepton
or jets lie outside the acceptance region.
We veto any event with more than one charged lepton with $p_T\,>$ 10 GeV.
The primary charged lepton has larger transverse momentum in pair production
due to the relatively large $p_T$ of the top-quark.
Large contributions are also seen for QCD production of $WJJ$ which 
has a harder $p_T$ spectrum for the charged lepton.
The $tW$ associated production can contribute at a level of tens
percents of the signal cross sections.
We note for signal of top-quark production, both $tW^-$
and $\bar t W^+$ production can contribute as backgrounds.
In the latter case the primary charged
lepton comes directly from $W^+$ decay leading to a harder
$p_T$ spectrum.
Backgrounds due to $s$-channel production or diboson production are
small.
A typical feature can be seen from Table~\ref{tab:bk} is that in signal
region CMS-SB where the light jet is required to be forward, almost all
backgrounds are suppressed by a factor of ten at least.
The signal from $t$-channel production are less affected due to the
forward nature of the light jet. 
\begin{table}[h!]
\centering
\begin{tabular}{l|c|cc|cc}
\hline
 [GeV]/[pb]	&&\multicolumn{2}{c}{CMS-SA} & \multicolumn{2}{c}{CMS-SB}\tabularnewline
\cline{3-6}
	&& $<$ 100 GeV & $<$ 200 GeV & $<$ 100 GeV & $<$ 200 GeV \tabularnewline
\hline
\hline
	\multirow{2}{*}{$t\bar t$} &$\langle p_{T,l}\rangle $ & 52.2 & 59.8 &51.9 &59.1 \tabularnewline
	& $\sigma_{fid}$  & 4.42 & 4.93 &0.40 & 0.44\tabularnewline
\hline
	\multirow{2}{*}{$tW^-$($\bar t W^+$)} &$\langle p_{T,l}\rangle $  & 52.2 & 61.8 &52.5 &61.1  \tabularnewline
	& $\sigma_{fid}$  &0.33&0.38&0.019&0.021 \tabularnewline
\hline
	\multirow{2}{*}{$s$-channel $t$} &$\langle p_{T,l}\rangle $  & 47.6 & 50.6 &47.2 &49.4  \tabularnewline
	& $\sigma_{fid}$  & 0.044 & 0.046 &0.007 & 0.007 \tabularnewline
\hline
	\multirow{2}{*}{$s$-channel $\bar t$} &$\langle p_{T,l}\rangle $ & 47.7 & 50.3 &47.4 &49.1 \tabularnewline
	& $\sigma_{fid}$ & 0.030 & 0.031 &0.004 & 0.004  \tabularnewline
\hline
	\multirow{2}{*}{QCD $W^+JJ$} &$\langle p_{T,l}\rangle $ & 50.5 & 59.2 & 51.0 & 58.8  \tabularnewline
	& $\sigma_{fid}$  & 1.29 & 1.45 & 0.157 & 0.174 \tabularnewline
\hline
	\multirow{2}{*}{QCD $W^-JJ$} &$\langle p_{T,l}\rangle $ & 52.5 & 64.2 & 52.9 & 62.8  \tabularnewline
	& $\sigma_{fid}$  & 0.99 & 1.15 & 0.107 & 0.117 \tabularnewline
\hline
	\multirow{2}{*}{$W^+Z$} &$\langle p_{T,l}\rangle $ & 53.0 & 65.1 & 55.2 & 68.5  \tabularnewline
	& $\sigma_{fid}$  & 0.005 & 0.006 & 0.0008 & 0.0009 \tabularnewline
\hline
	\multirow{2}{*}{$W^-Z$} &$\langle p_{T,l}\rangle $ & 52.7 & 63.5 & 51.8 & 60.2  \tabularnewline
	& $\sigma_{fid}$  & 0.004 & 0.004 & 0.0005 & 0.0006 \tabularnewline
\hline
	\multirow{2}{*}{$t$-channel $t$} &$\langle p_{T,l}\rangle $ & 47.65 & 50.10 & 47.35 &  49.25   \tabularnewline
	& $\sigma_{fid}$  & 1.10 & 1.14 & 0.493 & 0.506 \tabularnewline
\hline
	\multirow{2}{*}{$t$-channel $\bar t$} &$\langle p_{T,l}\rangle $ &47.85&50.17&47.70&49.54   \tabularnewline
	& $\sigma_{fid}$ &0.674&0.696&0.250&0.257  \tabularnewline
\hline
\end{tabular}

\caption{
Average transverse momentum of the charged lepton
and fiducial cross section within the two fiducial regions,
for various background processes to $t$-channel top quark and anti-quark
production.
The top-quark pair production or top-quark associated production
with $W$ boson contribute equally to the two charge conjugate final
states.
\label{tab:bk}}
\end{table}

From Table~\ref{tab:bk} we find even in the region CMS-SB, the rate
of $t\bar t$ background can still reach the same level as the signal processes.
That can easily spoil the precision on measurement of the average $p_T$ for the signal
processes due to uncertainties on modeling of the $t\bar t$ background.
Further more, any backgrounds from top-quark production depend on the
top-quark mass as well, which will complicate the extraction of the top-quark mass.
One important observation is that both the $t\bar t$ and $t W$ backgrounds
contribute almost equally to signal processes of charged
lepton with positive and negative charges.
The charge asymmetry first enters at NLO for $t\bar t$ production and is small at
the LHC.
In case of $t W$ production the asymmetry vanishes even at NLO.
Thus one possibility is to measure the difference of lepton $p_T$
spectrums for positive and negative charges.
Dependence and associated uncertainties on modeling of the
$t\bar t$ and $t W$ backgrounds are minimized, though their statistical
fluctuations remain.
Sensitivity of the signal processes to top-quark mass and the theoretical uncertainties
are almost unchanged when taking differences of
spectrums with opposite charges.
That is because the differences of $t$-channel single top quark
and anti-quark production are mostly driven by different parton distributions
at the light-quark line.
At the end, the uncertainties due to modeling of QCD $WJJ$ background will be
dominant.
However, as mentioned earlier, a large fraction of $WJJ$ background arise from
production of charm quark, gluon or light quarks which are misidentified
as $b$-jets.
One can further reduce their impact by either imposing a tighter $b$-tagging
criteria or using data-driven methods.  

\section{Projection for (HL-)LHC}
\label{sec:pro}

We provide an estimation on precision of the top-quark mass measurement
can be achieved in the coming run of LHC and HL-LHC.
As explained earlier, the observable used is the average transverse
momentum of the charged lepton in the charge-weighted distribution,
\begin{eqnarray}\label{eq:ptobs}
	\langle p_T \rangle_{obs} &&\equiv 
	\frac{\int p_{T,l}[d\sigma^{l^+}-d\sigma^{l^-}]}
	{\sigma^{l^+}-\sigma^{l^-}} \nonumber \\
	&&=\frac{\int p_{T,l}[d\sigma_S^{l^+}-d\sigma_S^{l^-}+d\sigma_B^{l^+}-d\sigma_B^{l^-}]}
	{\sigma_S^{l^+}-\sigma_S^{l^-}+\sigma_B^{l^+}-\sigma_B^{l^-}} \nonumber \\
	&&=\langle p_T \rangle_S
	+\frac{r}{1+r}\left[\langle p_T \rangle_B-\langle p_T \rangle_S\right],
\end{eqnarray}
where in the second line we have rewritten the average $p_T$ in terms
of signal and background contributions.
$\langle p_T \rangle_{S(B)}$ are the average $p_T$ of the charged lepton in the
spectrum of signal(background) alone, 
\begin{eqnarray}\label{eq:ptobs}
	\langle p_T \rangle_{S(B)} &&\equiv 
	\frac{\int p_{T,l}[d\sigma_{S(B)}^{l^+}-d\sigma_{S(B)}^{l^-}]}
	{\sigma_{S(B)}^{l^+}-\sigma_{S(B)}^{l^-}},
\end{eqnarray}
and $r$ is the background to signal ratio,
\begin{equation}
	r\equiv \frac{\sigma_B^{l^+}-\sigma_B^{l^-}}{\sigma_S^{l^+}-\sigma_S^{l^-}}.
\end{equation}
We neglect backgrounds other than from top-quark pair production
and QCD production of $WJJ$, which are small according to Table~\ref{tab:bk}.
From Eq.~(\ref{eq:ptobs}) we can extract the average transverse momentum
of the signal $\langle p_T \rangle_S$ using the
measurement on $\langle p_T \rangle_{obs}$ and inputs of $r$ and
$\langle p_T \rangle_B$.
From our theory calculation we can arrive at a linear model
on dependence of the average $p_T$ on the top-quark mass, 
\begin{equation}\label{eq:pts}
	\langle p_T \rangle_{S}=p_{T,0}+\lambda \left[\frac{m_t}{\rm GeV}-172.5\right],
\end{equation}
where $p_{T,0}$ is the average $p_T$ of signal for a top-quark mass of 172.5 GeV.
$p_{T,0}$ and $\lambda$ can be derived from NNLO predictions shown
in Tables~\ref{tab:5fpt}-\ref{tab:par} together with the counterparts for
top anti-quark production.
By combining Eqs.~(\ref{eq:ptobs}) and~(\ref{eq:pts}) we can extract the
top-quark mass.

\begin{figure}[ht]
\centering
  \includegraphics[width=0.8\textwidth]{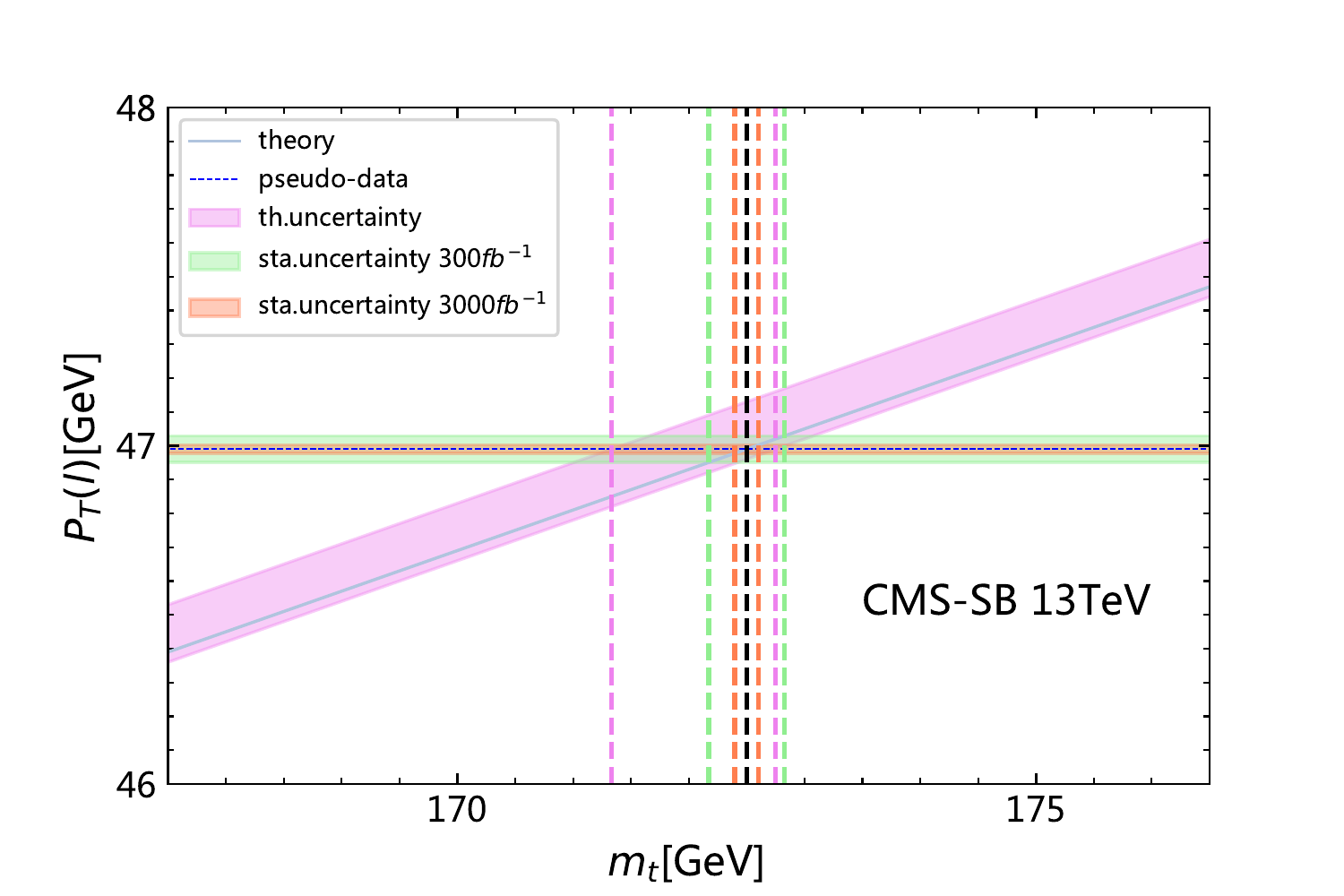}
\caption{
Predictions on the average transverse momentum of the charged lepton in
the signal process as a function of the top-quark mass (band
along diagonal direction) and the projected measurement on the same
quantity with only statistical errors (horizontal bands).
Extracted top-quark mass with various uncertainties are indicated
by vertical lines.
\label{fig:prjA}}
\end{figure}

\begin{figure}[ht]
\centering
  \includegraphics[width=0.8\textwidth]{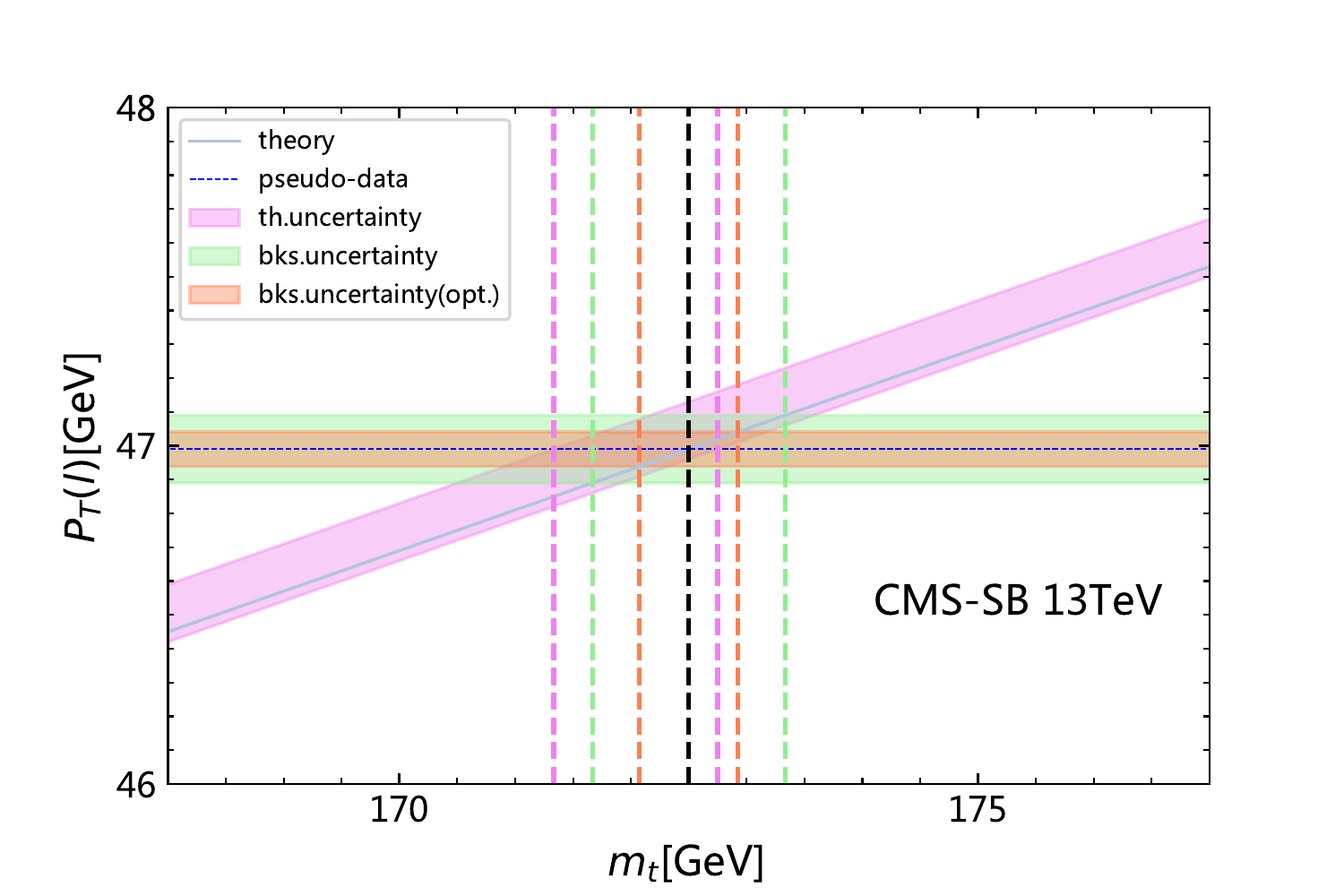}
\caption{
Predictions on the average transverse momentum of the charged lepton of
the signal process as a function of the top-quark mass (band
along diagonal direction) and the projected measurement on the same
quantity with only systematic errors from background modeling (horizontal bands).
Extracted top-quark mass with various uncertainties are indicated
by vertical lines.
\label{fig:prjB}}
\end{figure}

In the following we focus on the signal region CMS-SB with $p_{T,l}<100\,{\rm GeV}$.
It benefits from both lower backgrounds and smaller theoretical uncertainties.
We estimate several contributions to the final uncertainty of measured top-quark mass.
The statistical uncertainty on $\langle p_T \rangle_{obs}$ due to fluctuations of both signal
and backgrounds, including $t\bar t$ contributions, are computed
with pseudo experiment assuming an integrated luminosity of 300 and 3000~fb$^{-1}$
respectively and assuming top-quark decays into two families of leptons.
Theoretical uncertainties on $p_{T,0}$ are estimated with scale variations
of the NNLO predictions shown in Table~\ref{tab:5fpt}.
In Fig.~\ref{fig:prjA} we plot results on determination of the top-quark
mass with a hypothetical value of 172.5 GeV.
The horizontal bands indicate the statistical uncertainties as propagated
into $\langle p_T \rangle_{S}$.
The diagonal band represents the theory prediction of $\langle p_T \rangle_{S}$
as a function of the top-quark mass including scale variations. 
The projected uncertainties on the extracted top-quark mass are computed
assuming linear error propagation, and are represented by vertical lines.
For example, the statistical uncertainty is about $\pm 0.3(0.1)$~GeV with
an integrated luminosity of 300(3000)~fb$^{-1}$.
The theoretical uncertainty amounts to $+0.3$ and $-1.2$~GeV.
Further uncertainties are related to modeling of the backgrounds.
We only need to consider the $WJJ$ background in this case since the
systematic uncertainty for $t\bar t$ production cancels in
the charge weighted $p_T$ distribution.
A precise study on QCD $WJJ$ background is beyond the scope of
current paper, and can be carried out with dedicated MC simulations.
We simply assign empirical numbers on systematic uncertainties
of $\langle p_T \rangle_{B}$ and $r$ from $WJJ$ background.
On one hand we assume they are 0.5~GeV and 10\% respectively, and
reduced by a factor of two in the optimistic case.
The results are shown in Fig.~\ref{fig:prjB} with the horizontal
bands representing uncertainty of $\langle p_T \rangle_{S}$ as propagated
from systematic errors of backgrounds.
The uncertainty on measured top-quark mass is 0.8 and 0.4~GeV for the
two scenarios respectively as shown by vertical lines in Fig.~\ref{fig:prjB},
comparing to the theoretical uncertainty shown earlier.
Thus we expect the full error budget of the extracted top-quark
mass consists of a theoretical uncertainty of about 1~GeV from
signal modeling, a systematic uncertainty of 0.4~GeV due to background modeling,
and a much smaller statistical uncertainty.

\section{Summary}
\label{sec:sum}
In summary we have studied the determination of the top-quark mass
using leptonic observables in $t$-channel single top-quark production
at the LHC.
Extraction of the top-quark mass from single top-quark production
benefits from the fact that systematic uncertainties are partially uncorrelated to those
in top-quark pair production on both experimental and theory sides.
We demonstrate sensitivity of the average transverse momentum of the charged
lepton to the top-quark mass.
Leptonic observables are generally believed to be less affected
by various non-perturbative QCD effects and the jet energy scale uncertainties.
We identify an appropriate signal region for such a measurement at
the LHC with enhanced signal to background ratio as well as stable
theory predictions.

We present our NNLO QCD predictions under narrow width approximation
using structure function approach.
We show that QCD corrections in top-quark decay play important role
for such leptonic observables.
We find a good convergence on predictions of the average transverse momentum
of the charged lepton with scale uncertainties well under control.
By comparing our fixed-order predictions to predictions from MC generators we
find the parton shower resummation can capture part of the NLO and NNLO
corrections, and the hadronization effects are in general
small for leptonic observables.
Besides, we point out several corrections that need to be included when
comparing our NNLO predictions with data, including non-resonant corrections,
non-factorized QCD corrections, EW corrections, and so on.

Moreover, we estimate various SM backgrounds to the signals
considered.
We propose to use the charge weighted distribution in the measurement,
i.e., difference between distributions of charged lepton with positive
and negative electric charges.
That can reduce uncertainties due to modeling of SM backgrounds
which contribute equally to final states with different
charges, for example, backgrounds from QCD jets production, top-quark
pair production, and top-quark associated production with a W boson.
We construct a simple model on dependence of
the observed average transverse momentum of the charged lepton to the top
quark mass, and present projections for future (HL-)LHC measurement on top quark mass.
The statistical uncertainties and theoretical uncertainties due to hadronization
corrections are found to be small.
Scale variations in our signal modeling transfer into an uncertainty
of about 1~GeV on the extracted top-quark mass.
However, the scale variations should be considered as an optimistic estimation
on the uncertainty due to the missing higher-order corrections.
Future works on the unknown non-factorized corrections as well as on matching
NNLO calculations with parton showers can provide a better understanding of
the perturbative uncertainties.
Lastly theoretical uncertainty due to modeling of remaining SM backgrounds
is estimated to be 0.4$\sim$0.8~GeV.

\begin{acknowledgments}
The work of J.~Gao was sponsored by the National Natural
Science Foundation of China under the Grant No. 11875189 and No.11835005.
The authors would like to thank Kai Yan and Dingyu Shao
for proofreading of the manuscript.
\end{acknowledgments}

\bibliography{tmass}

\providecommand{\href}[2]{#2}\begingroup\raggedright\begin{thebibliography}{100}

\bibitem{1803.01853}
J.~Haller, A.~Hoecker, R.~Kogler, K.~Mönig, T.~Peiffer, and J.~Stelzer, {\it
  {Update of the global electroweak fit and constraints on two-Higgs-doublet
  models}},  {\em Eur. Phys. J.} {\bf C78} (2018), no.~8 675,
  [\href{http://arxiv.org/abs/1803.01853}{{\tt arXiv:1803.01853}}].

\bibitem{hep-ph/0104016}
G.~Isidori, G.~Ridolfi, and A.~Strumia, {\it {On the metastability of the
  standard model vacuum}},  {\em Nucl. Phys.} {\bf B609} (2001) 387--409,
  [\href{http://arxiv.org/abs/hep-ph/0104016}{{\tt hep-ph/0104016}}].

\bibitem{1407.2682}
{\bf CDF, D0} Collaboration, T.~E.~W. Group, {\it {Combination of CDF and D0
  Results on the Mass of the Top Quark using up to 9.7 fb$^{-1}$ at the
  Tevatron}},  \href{http://arxiv.org/abs/1407.2682}{{\tt arXiv:1407.2682}}.

\bibitem{1810.01772}
{\bf ATLAS} Collaboration, M.~Aaboud et~al., {\it {Measurement of the top quark
  mass in the $t\bar{t}\rightarrow $ lepton+jets channel from $\sqrt{s}=8$ TeV
  ATLAS data and combination with previous results}},  {\em Eur. Phys. J.} {\bf
  C79} (2019), no.~4 290, [\href{http://arxiv.org/abs/1810.01772}{{\tt
  arXiv:1810.01772}}].

\bibitem{1812.10534}
{\bf CMS} Collaboration, A.~M. Sirunyan et~al., {\it {Measurement of the top
  quark mass in the all-jets final state at $\sqrt{s} =$ 13 TeV and combination
  with the lepton+jets channel}},  {\em Eur. Phys. J.} {\bf C79} (2019), no.~4
  313, [\href{http://arxiv.org/abs/1812.10534}{{\tt arXiv:1812.10534}}].

\bibitem{1412.3649}
A.~H. Hoang, {\it {The Top Mass: Interpretation and Theoretical
  Uncertainties}},  in {\em {Proceedings, 7th International Workshop on Top
  Quark Physics (TOP2014): Cannes, France, September 28-October 3, 2014}},
  2014.
\newblock \href{http://arxiv.org/abs/1412.3649}{{\tt arXiv:1412.3649}}.

\bibitem{1712.02796}
P.~Nason, {\it {The Top Mass in Hadronic Collisions}},  in {\em From My Vast
  Repertoire ...: Guido Altarelli's Legacy} (A.~Levy, S.~Forte, and G.~Ridolfi,
  eds.), pp.~123--151.
\newblock 2019.
\newblock \href{http://arxiv.org/abs/1712.02796}{{\tt arXiv:1712.02796}}.

\bibitem{1807.06617}
A.~H. Hoang, S.~Plätzer, and D.~Samitz, {\it {On the Cutoff Dependence of the
  Quark Mass Parameter in Angular Ordered Parton Showers}},  {\em JHEP} {\bf
  10} (2018) 200, [\href{http://arxiv.org/abs/1807.06617}{{\tt
  arXiv:1807.06617}}].

\bibitem{1902.05035}
S.~Ferrario~Ravasio, {\em {Top-mass observables: all-orders behaviour,
  renormalons and NLO + Parton Shower effects}}.
\newblock PhD thesis, Milan Bicocca U., 2018.
\newblock \href{http://arxiv.org/abs/1902.05035}{{\tt arXiv:1902.05035}}.

\bibitem{FerrarioRavasio:2019vmq}
S.~Ferrario~Ravasio, T.~Ježo, P.~Nason, and C.~Oleari, {\it {A theoretical
  study of top-mass measurements at the LHC using NLO+PS generators of
  increasing accuracy}},  {\em Eur. Phys. J.} {\bf C78} (2018), no.~6 458,
  [\href{http://arxiv.org/abs/1906.09166}{{\tt arXiv:1906.09166}}]. [Addendum:
  Eur. Phys. J.C79,no.10,859(2019)].

\bibitem{1605.03609}
M.~Beneke, P.~Marquard, P.~Nason, and M.~Steinhauser, {\it {On the ultimate
  uncertainty of the top quark pole mass}},  {\em Phys. Lett.} {\bf B775}
  (2017) 63--70, [\href{http://arxiv.org/abs/1605.03609}{{\tt
  arXiv:1605.03609}}].

\bibitem{1704.01580}
A.~H. Hoang, A.~Jain, C.~Lepenik, V.~Mateu, M.~Preisser, I.~Scimemi, and I.~W.
  Stewart, {\it {The MSR mass and the $
  \mathcal{O}\left({\Lambda}_{\mathrm{QCD}}\right) $ renormalon sum rule}},
  {\em JHEP} {\bf 04} (2018) 003, [\href{http://arxiv.org/abs/1704.01580}{{\tt
  arXiv:1704.01580}}].

\bibitem{1706.08526}
A.~H. Hoang, C.~Lepenik, and M.~Preisser, {\it {On the Light Massive Flavor
  Dependence of the Large Order Asymptotic Behavior and the Ambiguity of the
  Pole Mass}},  {\em JHEP} {\bf 09} (2017) 099,
  [\href{http://arxiv.org/abs/1706.08526}{{\tt arXiv:1706.08526}}].

\bibitem{hep-ph/9906349}
C.~G. Lester and D.~J. Summers, {\it {Measuring masses of semiinvisibly
  decaying particles pair produced at hadron colliders}},  {\em Phys. Lett.}
  {\bf B463} (1999) 99--103, [\href{http://arxiv.org/abs/hep-ph/9906349}{{\tt
  hep-ph/9906349}}].

\bibitem{1407.2763}
S.~Frixione and A.~Mitov, {\it {Determination of the top quark mass from
  leptonic observables}},  {\em JHEP} {\bf 09} (2014) 012,
  [\href{http://arxiv.org/abs/1407.2763}{{\tt arXiv:1407.2763}}].

\bibitem{1603.03445}
K.~Agashe, R.~Franceschini, D.~Kim, and M.~Schulze, {\it {Top quark mass
  determination from the energy peaks of b-jets and B-hadrons at NLO QCD}},
  {\em Eur. Phys. J.} {\bf C76} (2016), no.~11 636,
  [\href{http://arxiv.org/abs/1603.03445}{{\tt arXiv:1603.03445}}].

\bibitem{1603.06536}
{\bf CMS} Collaboration, V.~Khachatryan et~al., {\it {Measurement of the top
  quark mass using charged particles in pp collisions at $\sqrt s =$ 8 TeV}},
  {\em Phys. Rev.} {\bf D93} (2016), no.~9 092006,
  [\href{http://arxiv.org/abs/1603.06536}{{\tt arXiv:1603.06536}}].

\bibitem{1608.03560}
{\bf CMS} Collaboration, V.~Khachatryan et~al., {\it {Measurement of the mass
  of the top quark in decays with a $J/\psi$ meson in pp collisions at 8 TeV}},
   {\em JHEP} {\bf 12} (2016) 123, [\href{http://arxiv.org/abs/1608.03560}{{\tt
  arXiv:1608.03560}}].

\bibitem{1709.09407}
{\bf ATLAS} Collaboration, M.~Aaboud et~al., {\it {Measurement of lepton
  differential distributions and the top quark mass in $t\bar{t}$ production in
  $pp$ collisions at $\sqrt{s}=8$ TeV with the ATLAS detector}},  {\em Eur.
  Phys. J.} {\bf C77} (2017), no.~11 804,
  [\href{http://arxiv.org/abs/1709.09407}{{\tt arXiv:1709.09407}}].

\bibitem{1406.5375}
{\bf ATLAS} Collaboration, G.~Aad et~al., {\it {Measurement of the $t\bar{t}$
  production cross-section using $e\mu $ events with b-tagged jets in pp
  collisions at $\sqrt{s}$ = 7 and 8 $\,\mathrm{TeV}$ with the ATLAS
  detector}},  {\em Eur. Phys. J.} {\bf C74} (2014), no.~10 3109,
  [\href{http://arxiv.org/abs/1406.5375}{{\tt arXiv:1406.5375}}]. [Addendum:
  Eur. Phys. J.C76,no.11,642(2016)].

\bibitem{1603.02303}
{\bf CMS} Collaboration, V.~Khachatryan et~al., {\it {Measurement of the t-tbar
  production cross section in the e-mu channel in proton-proton collisions at
  sqrt(s) = 7 and 8 TeV}},  {\em JHEP} {\bf 08} (2016) 029,
  [\href{http://arxiv.org/abs/1603.02303}{{\tt arXiv:1603.02303}}].

\bibitem{1812.10505}
{\bf CMS} Collaboration, A.~M. Sirunyan et~al., {\it {Measurement of the
  $\mathrm{t}\overline{\mathrm{t}}$ production cross section, the top quark
  mass, and the strong coupling constant using dilepton events in pp collisions
  at $\sqrt{s} =$ 13 TeV}},  {\em Eur. Phys. J.} {\bf C79} (2019), no.~5 368,
  [\href{http://arxiv.org/abs/1812.10505}{{\tt arXiv:1812.10505}}].

\bibitem{1904.05237}
{\bf CMS} Collaboration, A.~M. Sirunyan et~al., {\it {Measurement of
  $\mathrm{t\bar t}$ normalised multi-differential cross sections in pp
  collisions at $\sqrt s=13$ TeV, and simultaneous determination of the strong
  coupling strength, top quark pole mass, and parton distribution functions}},
  \href{http://arxiv.org/abs/1904.05237}{{\tt arXiv:1904.05237}}.

\bibitem{1908.02179}
W.-L. Ju, G.~Wang, X.~Wang, X.~Xu, Y.~Xu, and L.~L. Yang, {\it {Invariant-mass
  distribution of top-quark pairs and top-quark mass determination}},
  \href{http://arxiv.org/abs/1908.02179}{{\tt arXiv:1908.02179}}.

\bibitem{2004.03088}
W.-L. Ju, G.~Wang, X.~Wang, X.~Xu, Y.~Xu, and L.~L. Yang, {\it {Top quark pair
  production near threshold: single/double distributions and mass
  determination}},  {\em JHEP} {\bf 06} (2020) 158,
  [\href{http://arxiv.org/abs/2004.03088}{{\tt arXiv:2004.03088}}].

\bibitem{1303.6415}
S.~Alioli, P.~Fernandez, J.~Fuster, A.~Irles, S.-O. Moch, P.~Uwer, and M.~Vos,
  {\it {A new observable to measure the top-quark mass at hadron colliders}},
  {\em Eur. Phys. J.} {\bf C73} (2013) 2438,
  [\href{http://arxiv.org/abs/1303.6415}{{\tt arXiv:1303.6415}}].

\bibitem{ATLAS:2014baa}
{\bf ATLAS} Collaboration, T.~A. collaboration, {\it {Measurement of the top
  quark mass in topologies enhanced with single top-quarks produced in the
  t-channel in $\sqrt{s}=8\,\mathrm{TeV}$ ATLAS data}}, .

\bibitem{1703.02530}
{\bf CMS} Collaboration, A.~M. Sirunyan et~al., {\it {Measurement of the top
  quark mass using single top quark events in proton-proton collisions at
  $\sqrt{s}= 8$ TeV}},  {\em Eur. Phys. J.} {\bf C77} (2017), no.~5 354,
  [\href{http://arxiv.org/abs/1703.02530}{{\tt arXiv:1703.02530}}].

\bibitem{1608.05212}
S.~Alekhin, S.~Moch, and S.~Thier, {\it {Determination of the top-quark mass
  from hadro-production of single top-quarks}},  {\em Phys. Lett.} {\bf B763}
  (2016) 341--346, [\href{http://arxiv.org/abs/1608.05212}{{\tt
  arXiv:1608.05212}}].

\bibitem{1506.06864}
M.~Beneke, Y.~Kiyo, P.~Marquard, A.~Penin, J.~Piclum, and M.~Steinhauser, {\it
  {Next-to-Next-to-Next-to-Leading Order QCD Prediction for the Top Antitop
  $S$-Wave Pair Production Cross Section Near Threshold in $e^+e^-$
  Annihilation}},  {\em Phys. Rev. Lett.} {\bf 115} (2015), no.~19 192001,
  [\href{http://arxiv.org/abs/1506.06864}{{\tt arXiv:1506.06864}}].

\bibitem{2004.12915}
A.~H. Hoang, {\it {What is the Top Quark Mass?}},
  \href{http://arxiv.org/abs/2004.12915}{{\tt arXiv:2004.12915}}.

\bibitem{NUPHA.B435.23}
G.~Bordes and B.~van Eijk, {\it {Calculating QCD corrections to single top
  production in hadronic interactions}},  {\em Nucl. Phys.} {\bf B435} (1995)
  23--58.

\bibitem{hep-ph/9603265}
R.~Pittau, {\it {Final state QCD corrections to off-shell single top production
  in hadron collisions}},  {\em Phys. Lett.} {\bf B386} (1996) 397--402,
  [\href{http://arxiv.org/abs/hep-ph/9603265}{{\tt hep-ph/9603265}}].

\bibitem{hep-ph/9705398}
T.~Stelzer, Z.~Sullivan, and S.~Willenbrock, {\it {Single top quark production
  via $W$ - gluon fusion at next-to-leading order}},  {\em Phys. Rev.} {\bf
  D56} (1997) 5919--5927, [\href{http://arxiv.org/abs/hep-ph/9705398}{{\tt
  hep-ph/9705398}}].

\bibitem{hep-ph/9807340}
T.~Stelzer, Z.~Sullivan, and S.~Willenbrock, {\it {Single top quark production
  at hadron colliders}},  {\em Phys. Rev.} {\bf D58} (1998) 094021,
  [\href{http://arxiv.org/abs/hep-ph/9807340}{{\tt hep-ph/9807340}}].

\bibitem{hep-ph/0102126}
B.~W. Harris, E.~Laenen, L.~Phaf, Z.~Sullivan, and S.~Weinzierl, {\it {Fully
  differential QCD corrections to single top quark final states}},  {\em Int.
  J. Mod. Phys.} {\bf A16S1A} (2001) 379--381,
  [\href{http://arxiv.org/abs/hep-ph/0102126}{{\tt hep-ph/0102126}}].

\bibitem{hep-ph/0207055}
B.~W. Harris, E.~Laenen, L.~Phaf, Z.~Sullivan, and S.~Weinzierl, {\it {The
  Fully Differential Single Top Quark Cross-Section in Next to Leading Order
  QCD}},  {\em Phys. Rev.} {\bf D66} (2002) 054024,
  [\href{http://arxiv.org/abs/hep-ph/0207055}{{\tt hep-ph/0207055}}].

\bibitem{Sullivan:2004ie}
Z.~Sullivan, {\it {Understanding single-top-quark production and jets at hadron
  colliders}},  {\em Phys. Rev.} {\bf D70} (2004) 114012,
  [\href{http://arxiv.org/abs/hep-ph/0408049}{{\tt hep-ph/0408049}}].

\bibitem{hep-ph/0408158}
J.~M. Campbell, R.~K. Ellis, and F.~Tramontano, {\it {Single top production and
  decay at next-to-leading order}},  {\em Phys. Rev.} {\bf D70} (2004) 094012,
  [\href{http://arxiv.org/abs/hep-ph/0408158}{{\tt hep-ph/0408158}}].

\bibitem{hep-ph/0510224}
Z.~Sullivan, {\it {Angular correlations in single-top-quark and Wjj production
  at next-to-leading order}},  {\em Phys. Rev.} {\bf D72} (2005) 094034,
  [\href{http://arxiv.org/abs/hep-ph/0510224}{{\tt hep-ph/0510224}}].

\bibitem{hep-ph/0504230}
Q.-H. Cao, R.~Schwienhorst, J.~A. Benitez, R.~Brock, and C.~P. Yuan, {\it
  {Next-to-leading order corrections to single top quark production and decay
  at the Tevatron: 2. $t^-$ channel process}},  {\em Phys. Rev.} {\bf D72}
  (2005) 094027, [\href{http://arxiv.org/abs/hep-ph/0504230}{{\tt
  hep-ph/0504230}}].

\bibitem{1007.0893}
P.~Falgari, P.~Mellor, and A.~Signer, {\it {Production-decay interferences at
  NLO in QCD for $t$-channel single-top production}},  {\em Phys. Rev.} {\bf
  D82} (2010) 054028, [\href{http://arxiv.org/abs/1007.0893}{{\tt
  arXiv:1007.0893}}].

\bibitem{1012.5132}
R.~Schwienhorst, C.~P. Yuan, C.~Mueller, and Q.-H. Cao, {\it {Single top quark
  production and decay in the $t$-channel at next-to-leading order at the
  LHC}},  {\em Phys. Rev.} {\bf D83} (2011) 034019,
  [\href{http://arxiv.org/abs/1012.5132}{{\tt arXiv:1012.5132}}].

\bibitem{1102.5267}
P.~Falgari, F.~Giannuzzi, P.~Mellor, and A.~Signer, {\it {Off-shell effects for
  t-channel and s-channel single-top production at NLO in QCD}},  {\em Phys.
  Rev.} {\bf D83} (2011) 094013, [\href{http://arxiv.org/abs/1102.5267}{{\tt
  arXiv:1102.5267}}].

\bibitem{1305.7088}
A.~S. Papanastasiou, R.~Frederix, S.~Frixione, V.~Hirschi, and F.~Maltoni, {\it
  {Single-top $t$-channel production with off-shell and non-resonant effects}},
   {\em Phys. Lett.} {\bf B726} (2013) 223--227,
  [\href{http://arxiv.org/abs/1305.7088}{{\tt arXiv:1305.7088}}].

\bibitem{1406.4403}
P.~Kant, O.~M. Kind, T.~Kintscher, T.~Lohse, T.~Martini, S.~Mölbitz, P.~Rieck,
  and P.~Uwer, {\it {HatHor for single top-quark production: Updated
  predictions and uncertainty estimates for single top-quark production in
  hadronic collisions}},  {\em Comput. Phys. Commun.} {\bf 191} (2015) 74--89,
  [\href{http://arxiv.org/abs/1406.4403}{{\tt arXiv:1406.4403}}].

\bibitem{Carrazza:2018mix}
S.~Carrazza, R.~Frederix, K.~Hamilton, and G.~Zanderighi, {\it {MINLO t-channel
  single-top plus jet}},  {\em JHEP} {\bf 09} (2018) 108,
  [\href{http://arxiv.org/abs/1805.09855}{{\tt arXiv:1805.09855}}].

\bibitem{0903.0005}
J.~M. Campbell, R.~Frederix, F.~Maltoni, and F.~Tramontano, {\it
  {Next-to-Leading-Order Predictions for t-Channel Single-Top Production at
  Hadron Colliders}},  {\em Phys. Rev. Lett.} {\bf 102} (2009) 182003,
  [\href{http://arxiv.org/abs/0903.0005}{{\tt arXiv:0903.0005}}].

\bibitem{1603.01178}
R.~Frederix, S.~Frixione, A.~S. Papanastasiou, S.~Prestel, and P.~Torrielli,
  {\it {Off-shell single-top production at NLO matched to parton showers}},
  {\em JHEP} {\bf 06} (2016) 027, [\href{http://arxiv.org/abs/1603.01178}{{\tt
  arXiv:1603.01178}}].

\bibitem{Neumann:2019kvk}
T.~Neumann and Z.~E. Sullivan, {\it {Off-Shell Single-Top-Quark Production in
  the Standard Model Effective Field Theory}},  {\em JHEP} {\bf 06} (2019) 022,
  [\href{http://arxiv.org/abs/1903.11023}{{\tt arXiv:1903.11023}}].

\bibitem{1907.12586}
R.~Frederix, D.~Pagani, and I.~Tsinikos, {\it {Precise predictions for
  single-top production: the impact of EW corrections and QCD shower on the
  $t$-channel signature}},  {\em JHEP} {\bf 09} (2019) 122,
  [\href{http://arxiv.org/abs/1907.12586}{{\tt arXiv:1907.12586}}].

\bibitem{1010.4509}
J.~Wang, C.~S. Li, H.~X. Zhu, and J.~J. Zhang, {\it {Factorization and
  resummation of t-channel single top quark production}},
  \href{http://arxiv.org/abs/1010.4509}{{\tt arXiv:1010.4509}}.

\bibitem{1103.2792}
N.~Kidonakis, {\it {Next-to-next-to-leading-order collinear and soft gluon
  corrections for t-channel single top quark production}},  {\em Phys. Rev.}
  {\bf D83} (2011) 091503, [\href{http://arxiv.org/abs/1103.2792}{{\tt
  arXiv:1103.2792}}].

\bibitem{1210.7698}
J.~Wang, C.~S. Li, and H.~X. Zhu, {\it {Resummation prediction on top quark
  transverse momentum distribution at large $p_T$}},  {\em Phys. Rev.} {\bf
  D87} (2013), no.~3 034030, [\href{http://arxiv.org/abs/1210.7698}{{\tt
  arXiv:1210.7698}}].

\bibitem{1510.06361}
N.~Kidonakis, {\it {Single-top transverse-momentum distributions at approximate
  NNLO}},  {\em Phys. Rev.} {\bf D93} (2016), no.~5 054022,
  [\href{http://arxiv.org/abs/1510.06361}{{\tt arXiv:1510.06361}}].

\bibitem{1801.09656}
Q.-H. Cao, P.~Sun, B.~Yan, C.~P. Yuan, and F.~Yuan, {\it {Transverse Momentum
  Resummation for $t$-channel single top quark production at the LHC}},  {\em
  Phys. Rev.} {\bf D98} (2018), no.~5 054032,
  [\href{http://arxiv.org/abs/1801.09656}{{\tt arXiv:1801.09656}}].

\bibitem{Kidonakis:2019nqa}
N.~Kidonakis, {\it {Soft anomalous dimensions for single-top production at
  three loops}},  {\em Phys. Rev.} {\bf D99} (2019), no.~7 074024,
  [\href{http://arxiv.org/abs/1901.09928}{{\tt arXiv:1901.09928}}].

\bibitem{Cao:2019uor}
Q.-H. Cao, P.~Sun, B.~Yan, C.~P. Yuan, and F.~Yuan, {\it {Soft Gluon
  Resummation in $t$-channel single top quark production at the LHC}},
  \href{http://arxiv.org/abs/1902.09336}{{\tt arXiv:1902.09336}}.

\bibitem{hep-ph/0512250}
S.~Frixione, E.~Laenen, P.~Motylinski, and B.~R. Webber, {\it {Single-top
  production in MC@NLO}},  {\em JHEP} {\bf 03} (2006) 092,
  [\href{http://arxiv.org/abs/hep-ph/0512250}{{\tt hep-ph/0512250}}].

\bibitem{0907.4076}
S.~Alioli, P.~Nason, C.~Oleari, and E.~Re, {\it {NLO single-top production
  matched with shower in POWHEG: s- and t-channel contributions}},  {\em JHEP}
  {\bf 09} (2009) 111, [\href{http://arxiv.org/abs/0907.4076}{{\tt
  arXiv:0907.4076}}]. [Erratum: JHEP02,011(2010)].

\bibitem{1207.5391}
R.~Frederix, E.~Re, and P.~Torrielli, {\it {Single-top t-channel
  hadroproduction in the four-flavour scheme with POWHEG and aMC@NLO}},  {\em
  JHEP} {\bf 09} (2012) 130, [\href{http://arxiv.org/abs/1207.5391}{{\tt
  arXiv:1207.5391}}].

\bibitem{1404.7116}
M.~Brucherseifer, F.~Caola, and K.~Melnikov, {\it {On the NNLO QCD corrections
  to single-top production at the LHC}},  {\em Phys. Lett.} {\bf B736} (2014)
  58--63, [\href{http://arxiv.org/abs/1404.7116}{{\tt arXiv:1404.7116}}].

\bibitem{Berger:2016oht}
E.~L. Berger, J.~Gao, C.~P. Yuan, and H.~X. Zhu, {\it {NNLO QCD Corrections to
  t-channel Single Top-Quark Production and Decay}},  {\em Phys. Rev.} {\bf
  D94} (2016), no.~7 071501, [\href{http://arxiv.org/abs/1606.08463}{{\tt
  arXiv:1606.08463}}].

\bibitem{1708.09405}
E.~L. Berger, J.~Gao, and H.~X. Zhu, {\it {Differential Distributions for
  t-channel Single Top-Quark Production and Decay at Next-to-Next-to-Leading
  Order in QCD}},  {\em JHEP} {\bf 11} (2017) 158,
  [\href{http://arxiv.org/abs/1708.09405}{{\tt arXiv:1708.09405}}].

\bibitem{1807.03835}
Z.~L. Liu and J.~Gao, {\it {s -channel single top quark production and decay at
  next-to-next-to-leading-order in QCD}},  {\em Phys. Rev.} {\bf D98} (2018),
  no.~7 071501, [\href{http://arxiv.org/abs/1807.03835}{{\tt
  arXiv:1807.03835}}].

\bibitem{2005.12936}
J.~Gao and E.~L. Berger, {\it {Modeling of $t$-channel single top-quark
  production at the LHC}},  \href{http://arxiv.org/abs/2005.12936}{{\tt
  arXiv:2005.12936}}.

\bibitem{Lindfors:1985zz}
J.~Lindfors, {\it {Higgs Boson Production by $W$ and $Z$ Collisions}},  {\em
  Phys. Lett.} {\bf 167B} (1986) 471--475.

\bibitem{Han:1992hr}
T.~Han, G.~Valencia, and S.~Willenbrock, {\it {Structure function approach to
  vector boson scattering in p p collisions}},  {\em Phys. Rev. Lett.} {\bf 69}
  (1992) 3274--3277, [\href{http://arxiv.org/abs/hep-ph/9206246}{{\tt
  hep-ph/9206246}}].

\bibitem{Stelzer:1997ns}
T.~Stelzer, Z.~Sullivan, and S.~Willenbrock, {\it {Single top quark production
  via $W$ - gluon fusion at next-to-leading order}},  {\em Phys. Rev.} {\bf
  D56} (1997) 5919--5927, [\href{http://arxiv.org/abs/hep-ph/9705398}{{\tt
  hep-ph/9705398}}].

\bibitem{Assadsolimani:2014oga}
M.~Assadsolimani, P.~Kant, B.~Tausk, and P.~Uwer, {\it {Calculation of two-loop
  QCD corrections for hadronic single top-quark production in the $t$
  channel}},  {\em Phys. Rev.} {\bf D90} (2014), no.~11 114024,
  [\href{http://arxiv.org/abs/1409.3654}{{\tt arXiv:1409.3654}}].

\bibitem{Meyer:2016slj}
C.~Meyer, {\it {Transforming differential equations of multi-loop Feynman
  integrals into canonical form}},  {\em JHEP} {\bf 04} (2017) 006,
  [\href{http://arxiv.org/abs/1611.01087}{{\tt arXiv:1611.01087}}].

\bibitem{hep-ph/9402326}
A.~Czarnecki and M.~Jezabek, {\it {Distributions of leptons in decays of
  polarized heavy quarks}},  {\em Nucl. Phys.} {\bf B427} (1994) 3--21,
  [\href{http://arxiv.org/abs/hep-ph/9402326}{{\tt hep-ph/9402326}}].

\bibitem{Tanabashi:2018oca}
{\bf Particle Data Group} Collaboration, M.~Tanabashi et~al., {\it {Review of
  Particle Physics}},  {\em Phys. Rev.} {\bf D98} (2018), no.~3 030001.

\bibitem{1907.08330}
{\bf CMS} Collaboration, A.~M. Sirunyan et~al., {\it {Measurement of
  differential cross sections and charge ratios for t-channel single top quark
  production in proton–proton collisions at $\sqrt{s}=13\,\text {Te}\text
  {V}$}},  {\em Eur. Phys. J.} {\bf C80} (2020), no.~5 370,
  [\href{http://arxiv.org/abs/1907.08330}{{\tt arXiv:1907.08330}}].

\bibitem{0802.1189}
M.~Cacciari, G.~P. Salam, and G.~Soyez, {\it {The Anti-k(t) jet clustering
  algorithm}},  {\em JHEP} {\bf 04} (2008) 063,
  [\href{http://arxiv.org/abs/0802.1189}{{\tt arXiv:0802.1189}}].

\bibitem{Stewart:2010tn}
I.~W. Stewart, F.~J. Tackmann, and W.~J. Waalewijn, {\it {N-Jettiness: An
  Inclusive Event Shape to Veto Jets}},  {\em Phys. Rev. Lett.} {\bf 105}
  (2010) 092002, [\href{http://arxiv.org/abs/1004.2489}{{\tt
  arXiv:1004.2489}}].

\bibitem{Boughezal:2015dva}
R.~Boughezal, C.~Focke, X.~Liu, and F.~Petriello, {\it {$W$-boson production in
  association with a jet at next-to-next-to-leading order in perturbative
  QCD}},  {\em Phys. Rev. Lett.} {\bf 115} (2015), no.~6 062002,
  [\href{http://arxiv.org/abs/1504.02131}{{\tt arXiv:1504.02131}}].

\bibitem{Gaunt:2015pea}
J.~Gaunt, M.~Stahlhofen, F.~J. Tackmann, and J.~R. Walsh, {\it {N-jettiness
  Subtractions for NNLO QCD Calculations}},  {\em JHEP} {\bf 09} (2015) 058,
  [\href{http://arxiv.org/abs/1505.04794}{{\tt arXiv:1505.04794}}].

\bibitem{Berger:2016inr}
E.~L. Berger, J.~Gao, C.~S. Li, Z.~L. Liu, and H.~X. Zhu, {\it {Charm-Quark
  Production in Deep-Inelastic Neutrino Scattering at Next-to-Next-to-Leading
  Order in QCD}},  {\em Phys. Rev. Lett.} {\bf 116} (2016), no.~21 212002,
  [\href{http://arxiv.org/abs/1601.05430}{{\tt arXiv:1601.05430}}].

\bibitem{1506.02660}
M.~Cacciari, F.~A. Dreyer, A.~Karlberg, G.~P. Salam, and G.~Zanderighi, {\it
  {Fully Differential Vector-Boson-Fusion Higgs Production at
  Next-to-Next-to-Leading Order}},  {\em Phys. Rev. Lett.} {\bf 115} (2015),
  no.~8 082002, [\href{http://arxiv.org/abs/1506.02660}{{\tt
  arXiv:1506.02660}}]. [Erratum: Phys. Rev. Lett.120,no.13,139901(2018)].

\bibitem{Gao:2012ja}
J.~Gao, C.~S. Li, and H.~X. Zhu, {\it {Top Quark Decay at Next-to-Next-to
  Leading Order in QCD}},  {\em Phys. Rev. Lett.} {\bf 110} (2013), no.~4
  042001, [\href{http://arxiv.org/abs/1210.2808}{{\tt arXiv:1210.2808}}].

\bibitem{Butterworth:2015oua}
J.~Butterworth et~al., {\it {PDF4LHC recommendations for LHC Run II}},  {\em J.
  Phys.} {\bf G43} (2016) 023001, [\href{http://arxiv.org/abs/1510.03865}{{\tt
  arXiv:1510.03865}}].

\bibitem{Gao:2013bia}
J.~Gao and P.~Nadolsky, {\it {A meta-analysis of parton distribution
  functions}},  {\em JHEP} {\bf 07} (2014) 035,
  [\href{http://arxiv.org/abs/1401.0013}{{\tt arXiv:1401.0013}}].

\bibitem{Harland-Lang:2014zoa}
L.~A. Harland-Lang, A.~D. Martin, P.~Motylinski, and R.~S. Thorne, {\it {Parton
  distributions in the LHC era: MMHT 2014 PDFs}},  {\em Eur. Phys. J.} {\bf
  C75} (2015), no.~5 204, [\href{http://arxiv.org/abs/1412.3989}{{\tt
  arXiv:1412.3989}}].

\bibitem{Ball:2014uwa}
{\bf NNPDF} Collaboration, R.~D. Ball et~al., {\it {Parton distributions for
  the LHC Run II}},  {\em JHEP} {\bf 04} (2015) 040,
  [\href{http://arxiv.org/abs/1410.8849}{{\tt arXiv:1410.8849}}].

\bibitem{Dulat:2015mca}
S.~Dulat, T.-J. Hou, J.~Gao, M.~Guzzi, J.~Huston, P.~Nadolsky, J.~Pumplin,
  C.~Schmidt, D.~Stump, and C.~P. Yuan, {\it {New parton distribution functions
  from a global analysis of quantum chromodynamics}},  {\em Phys. Rev.} {\bf
  D93} (2016), no.~3 033006, [\href{http://arxiv.org/abs/1506.07443}{{\tt
  arXiv:1506.07443}}].

\bibitem{Carrazza:2015aoa}
S.~Carrazza, S.~Forte, Z.~Kassabov, J.~I. Latorre, and J.~Rojo, {\it {An
  Unbiased Hessian Representation for Monte Carlo PDFs}},  {\em Eur. Phys. J.}
  {\bf C75} (2015), no.~8 369, [\href{http://arxiv.org/abs/1505.06736}{{\tt
  arXiv:1505.06736}}].

\bibitem{1203.6393}
F.~Maltoni, G.~Ridolfi, and M.~Ubiali, {\it {b-initiated processes at the LHC:
  a reappraisal}},  {\em JHEP} {\bf 07} (2012) 022,
  [\href{http://arxiv.org/abs/1203.6393}{{\tt arXiv:1203.6393}}]. [Erratum:
  JHEP04,095(2013)].

\bibitem{Harland-Lang:2015qea}
L.~A. Harland-Lang, A.~D. Martin, P.~Motylinski, and R.~S. Thorne, {\it {Charm
  and beauty quark masses in the MMHT2014 global PDF analysis}},  {\em Eur.
  Phys. J.} {\bf C76} (2016), no.~1 10,
  [\href{http://arxiv.org/abs/1510.02332}{{\tt arXiv:1510.02332}}].

\bibitem{1711.02568}
E.~Bothmann, F.~Krauss, and M.~Schönherr, {\it {Single top-quark production
  with SHERPA}},  {\em Eur. Phys. J.} {\bf C78} (2018), no.~3 220,
  [\href{http://arxiv.org/abs/1711.02568}{{\tt arXiv:1711.02568}}].

\bibitem{Campbell:2016jau}
J.~M. Campbell, R.~K. Ellis, and C.~Williams, {\it {Associated production of a
  Higgs boson at NNLO}},  {\em JHEP} {\bf 06} (2016) 179,
  [\href{http://arxiv.org/abs/1601.00658}{{\tt arXiv:1601.00658}}].

\bibitem{Boughezal:2016wmq}
R.~Boughezal, J.~M. Campbell, R.~K. Ellis, C.~Focke, W.~Giele, X.~Liu,
  F.~Petriello, and C.~Williams, {\it {Color singlet production at NNLO in
  MCFM}},  {\em Eur. Phys. J.} {\bf C77} (2017), no.~1 7,
  [\href{http://arxiv.org/abs/1605.08011}{{\tt arXiv:1605.08011}}].

\bibitem{Alwall:2014hca}
J.~Alwall, R.~Frederix, S.~Frixione, V.~Hirschi, F.~Maltoni, O.~Mattelaer,
  H.~S. Shao, T.~Stelzer, P.~Torrielli, and M.~Zaro, {\it {The automated
  computation of tree-level and next-to-leading order differential cross
  sections, and their matching to parton shower simulations}},  {\em JHEP} {\bf
  07} (2014) 079, [\href{http://arxiv.org/abs/1405.0301}{{\tt
  arXiv:1405.0301}}].

\bibitem{Artoisenet:2012st}
P.~Artoisenet, R.~Frederix, O.~Mattelaer, and R.~Rietkerk, {\it {Automatic
  spin-entangled decays of heavy resonances in Monte Carlo simulations}},  {\em
  JHEP} {\bf 03} (2013) 015, [\href{http://arxiv.org/abs/1212.3460}{{\tt
  arXiv:1212.3460}}].

\bibitem{Sjostrand:2006za}
T.~Sjostrand, S.~Mrenna, and P.~Z. Skands, {\it {PYTHIA 6.4 Physics and
  Manual}},  {\em JHEP} {\bf 05} (2006) 026,
  [\href{http://arxiv.org/abs/hep-ph/0603175}{{\tt hep-ph/0603175}}].

\bibitem{Sjostrand:2014zea}
T.~Sjöstrand, S.~Ask, J.~R. Christiansen, R.~Corke, N.~Desai, P.~Ilten,
  S.~Mrenna, S.~Prestel, C.~O. Rasmussen, and P.~Z. Skands, {\it {An
  Introduction to PYTHIA 8.2}},  {\em Comput. Phys. Commun.} {\bf 191} (2015)
  159--177, [\href{http://arxiv.org/abs/1410.3012}{{\tt 1410.3012}}].

\bibitem{Bahr:2008pv}
M.~Bahr et~al., {\it {Herwig++ Physics and Manual}},  {\em Eur. Phys. J.} {\bf
  C58} (2008) 639--707, [\href{http://arxiv.org/abs/0803.0883}{{\tt
  arXiv:0803.0883}}].

\bibitem{Conte:2012fm}
E.~Conte, B.~Fuks, and G.~Serret, {\it {MadAnalysis 5, A User-Friendly
  Framework for Collider Phenomenology}},  {\em Comput. Phys. Commun.} {\bf
  184} (2013) 222--256, [\href{http://arxiv.org/abs/1206.1599}{{\tt
  arXiv:1206.1599}}].

\bibitem{Cacciari:2011ma}
M.~Cacciari, G.~P. Salam, and G.~Soyez, {\it {FastJet User Manual}},  {\em Eur.
  Phys. J.} {\bf C72} (2012) 1896, [\href{http://arxiv.org/abs/1111.6097}{{\tt
  1111.6097}}].

\bibitem{1607.04538}
T.~Ježo, J.~M. Lindert, P.~Nason, C.~Oleari, and S.~Pozzorini, {\it {An NLO+PS
  generator for $t\bar{t}$ and $Wt$ production and decay including non-resonant
  and interference effects}},  {\em Eur. Phys. J.} {\bf C76} (2016), no.~12
  691, [\href{http://arxiv.org/abs/1607.04538}{{\tt arXiv:1607.04538}}].

\bibitem{Liu:2019tuy}
T.~Liu, K.~Melnikov, and A.~A. Penin, {\it {Nonfactorizable QCD Effects in
  Higgs Boson Production via Vector Boson Fusion}},  {\em Phys. Rev. Lett.}
  {\bf 123} (2019), no.~12 122002, [\href{http://arxiv.org/abs/1906.10899}{{\tt
  arXiv:1906.10899}}].

\bibitem{Dreyer:2020urf}
F.~A. Dreyer, A.~Karlberg, and L.~Tancredi, {\it {On the impact of
  non-factorisable corrections in VBF single and double Higgs production}},
  \href{http://arxiv.org/abs/2005.11334}{{\tt arXiv:2005.11334}}.

\end{thebibliography}\endgroup
\bibliographystyle{jhep}

\end{document}